\documentclass[12pt, draftclsnofoot, onecolumn,comsoc]{IEEEtran}

\usepackage{float}
\usepackage{graphicx}
\usepackage{clrscode}
\usepackage{stfloats}
\usepackage{paralist}
\usepackage{cite}
\usepackage{comment}
\usepackage{graphicx}
\usepackage{amsmath, amssymb, mathrsfs, amsfonts, amsthm}
\usepackage{epsfig, epstopdf}
\usepackage[tight,footnotesize]{subfigure}
\usepackage{algorithm}
\usepackage{color}
\usepackage[usenames,dvipsnames,svgnames,table]{xcolor}
\usepackage{bbold}
                           
\usepackage{algorithmic}

\usepackage{tikz}
\usetikzlibrary{arrows,automata}
\usepackage{mathtools}
\usepackage{enumitem}
\usepackage{algorithmic}
\usepackage{hyperref}

\newtheorem{lemma}{Lemma}
\newtheorem{remark}{Remark}

\def\msP{ \mathsf{P}}

\newcommand{\HTemp}{\mathcal{H}}
\newcommand{\HH}{\mathcal{H}_0}
\newcommand{\HHH}{\mathcal{H}_1}

\newcommand{\Pf}{ \text{P}_{\text{fa}} }

\newcommand{\Pm}{ \text{P}_{\text{m}} }

\newcommand{\Po}{\mathsf{Poisson}}
\newcommand{\Ex}{\mathsf{exp}}

\newcommand{\J}{\zeta}
\newcommand{\A}{\vartheta}
\newcommand{\Db}{D_{\text{b}}}

\newcommand{\K}{k_{\text{Deg}}}
\newcommand{\CT}{{\mathsf{C}}}
\newcommand{\F}{{\mathsf{J}}}

\allowdisplaybreaks

\begin{document}

\title{Early Cancer Detection in Blood Vessels Using Mobile Nanosensors}

\author{ Reza Mosayebi}
\author{Reza Mosayebi$^\dagger$, Arman Ahmadzadeh$^{\ddagger}$, Wayan Wicke$^\ddagger$, Vahid Jamali$^{\ddagger}$, \\Robert Schober$^{\ddagger}$, and Masoumeh Nasiri-Kenari$^\dagger$ \\$^\dagger$Sharif University of Technology, Tehran, Iran, \\$^{\ddagger}$University of Erlangen-Nuremberg, Erlangen, Germany}
\maketitle

\vspace{-1.6cm}

\begin{abstract}
%
In this paper, we propose using mobile nanosensors (MNSs) for early stage anomaly detection. For concreteness, we focus on the detection of cancer cells located in a particular region of a blood vessel. These cancer cells produce and emit special molecules, so-called biomarkers, which are symptomatic for the presence of anomaly, into the cardiovascular system. 
Detection of cancer biomarkers with conventional blood tests is difficult in the early stages of a cancer due to the very low concentration of the biomarkers in the samples taken. However, close to the cancer cells, the concentration of the cancer biomarkers is high. Hence, detection is possible if a sensor with the ability to detect these biomarkers is placed in the vicinity of the cancer cells.
Therefore, in this paper, we study the use of MNSs that are injected at a suitable injection site and can move through the blood vessels of the cardiovascular system, which potentially contain cancer cells. These MNSs can be activated by the biomarkers close to the cancer cells, where the biomarker concentration is sufficiently high. Eventually, the MNSs are collected by a fusion center (FC) where their activation levels are read and exploited to declare the presence of anomaly. 
We analytically derive the biomarker concentration in cancerous blood vessels as well as the probability mass function of the MNSs' activation levels and validate the obtained results via particle-based simulations. Then, we derive the optimal decision rule for the FC regarding the presence of anomaly assuming that the entire network is known at the FC. Finally, for the FC, we propose a simple sum detector that does not require knowledge of the network topology. Our simulations reveal that while the LRT detector achieves a higher performance than the sum detector, both proposed detectors significantly outperform a benchmark scheme that uses fixed nanosensors at the FC.
\end{abstract}
%


\vspace*{-4.5mm}
\section{Introduction}
\vspace*{-0.5mm}

Molecular communication (MC) is an emerging technology enabling communication among nanomachines. Inspired by biological systems, synthetic diffusion-based MC systems have been proposed as a potential solution for communication in nanonetworks where molecules play the role of information carriers \cite{MC_Book}. Nanonetworks are envisioned to facilitate revolutionary applications in areas such as biological engineering, healthcare, and environmental monitoring~\cite{Akyl_MCNet, Farsad, Nakano}.

One of the key challenges in health monitoring and disease diagnosis applications is the problem of \emph{anomaly detection}, e.g., \emph{early cancer detection}, which has received significant attention in medicine and other related fields \cite{Alok_Mishra,LiWu}. Since early cancer detection can significantly decrease cancer mortality, great efforts have been devoted to the investigation of new technologies for detecting the symptoms of cancer at an early stage \cite{Alok_Mishra,LiWu, Keesee,Tarro,Hanash}. These symptoms are characteristics that can indicate the presence of anomaly, and include \emph{cancer biomarkers} \cite{Alok_Mishra}. 
Cancer biomarkers cover a broad range of biochemical entities such as nucleic acids, proteins, sugars, small metabolites, and cytogenetic and cytokinetic parameters as well as entire cancer cells found in body fluids \cite{LiWu}. 
Among these biomarkers, proteins are of particular interest since they are primarily found in blood and urine where they can be measured with current medical technologies, such as clinical blood tests\cite{Keesee, Tarro}. 
It has been shown in \cite{Hanash, Miroslava_Stanstna, Karagiannis} that abnormal behavior/expressions of protein biomarkers can be associated with particular cancers. For example, $\alpha$-fetoprotein, carcinoma antigen $125$, carcinoembryonic antigen, and prostate-specific antigen are common biomarkers for liver, ovarian, colorectal, and prostate cancers, respectively \cite{Henry}.

Conventional blood tests may not be able to detect biomarkers secreted by cancer cells in the early stages of a cancer due to the very low concentration of the biomarkers inside the cardiovascular system (CS) \cite{Keesee, Tarro}. However, close to the cancer cells, the concentration of the cancer biomarkers is high such that reliable detection is possible if a corresponding sensor passes in the vicinity of the cancer cells. 
In this paper, we propose the use of engineered nanosensors for this purpose. Such nanosensors play a key role in nanomedicine and can carry and deliver imaging probes, therapeutic agents, and biological materials to target sites such as specific organs, tissues, and even particular cells \cite{LiWu, Chen, Drug}. 
The ability of engineered nanosensors to fast and intelligently release, move, observe, and read inside the CS motivates the investigation of the use of mobile nanosensors (MNSs) for anomaly detection \cite{Perfezou}. In particular, MNSs can be released from an injection site, move through the CS, become activated at sites of high biomarker concentration, and eventually be captured at a fusion center (FC) which then decides on the presence of an anomaly. The time interval between release and capture of the MNSs at the FC is called the \emph{observation window} and is on the order of a few minutes due to the fast flow velocities inside CS.

Anomaly detection has been extensively studied in different fields, including computer science, segmentation of biomedical signals, and fraud detection for credit cards, see e.g. \cite{Survey_Anomaly} and \cite{Computer_Failor}. However, in the context of MC, the amount of related prior work is limited. 
In \cite{Lahouti_Detection}, anomaly detection in molecular nanonetworks is studied and a suboptimal decision rule is employed to combine the observations at the FC. More recently, in \cite{Reza_TNB}, both optimal and near-optimal decision rules are developed for networks where nanosensors employ either one or multiple types of molecules to relay their gathered information to the FC. However, both \cite{Lahouti_Detection} and \cite{Reza_TNB} assume that the nanosensors are \emph{fixed}, i.e., they do not move inside the CS. 
In \cite{Luca}, an MC system for tumor detection in blood vessels is proposed. In this work, the authors assume that specific nanorobots are injected into the CS which are attracted by the tumor cells. After detection of a tumor, these nanorobots remain close to the tumor site and release secondary nanomachines inside the blood vessel to relay the nanorobots' information to specific receivers. 
Finally, in \cite{Nakano_Let}, a graph-based model for mobile MC systems with several bio-nanomachines is proposed, and the concentration of the bio-nanomachines is numerically evaluated. It is shown that similar to \cite{Luca}, the concentration of the considered bio-nanomachines is high in the vicinity of tumor cells.

We note that the cardiovascular network is highly complex and the propagation of the molecules within this network is therefore very complicated. Hence, most prior works relied on extensive simulations to analyze sensing systems for the CS \cite{Luca, Nakano_Let}. On the contrary, in this paper, by developing a simplified yet meaningful model for the CS, we establish an analytical framework for analyzing the proposed detection system which provides useful insights for system design.
In particular, we consider collaborative anomaly detection where multiple MNSs are released at an injection site into the CS to individually detect biomarkers. The MNSs are transported through the CS and can detect the presence of biomarkers, which triggers the activation of the MNSs. Nanosensors with such detection capabilities have already been reported in the literature, see \cite{Chen} and references therein. 
The MNSs may ultimately reach an FC which decides on the presence of an anomaly based on the activation levels of the observed MNSs. Relying on multiple MNSs is motivated by the fact that, in general, healthy cells also release a small amount of the same type of biomarkers into the CS as the cancer cells. This introduces noise to the sensing process of the MNS, which may have a large impact in the early stages of a cancer and may lead to unreliable decisions. 
Moreover, a given MNS may not pass both the cancer cell or the FC on its path through the CS.
Therefore, detection reliability can be improved by employing multiple MNSs.
The contributions of this paper can be summarized as follows.
\begin{enumerate}
\item[] $\bullet$ Although in general the CS has a highly complex structure, we propose a simplified yet meaningful model for the blood vessels of the CS facilitating first-order insights about the propagation of biomarkers and MNSs through the CS. In particular, we model blood vessels as two-dimensional (2-D) rectangles and simplify existing models for the release rate of biomarkers for both healthy and cancerous cells. We note that, in the presence of cancer cells, the total number of biomarkers inside the blood vessels is a function of time. Nevertheless, since the change in production rate of cancer biomarkers is very slow \cite{Brown}, we assume a \emph{quasi-steady} state behavior for the number of biomarkers in the blood vessels during the comparatively short observation window. 
\item[] $\bullet$ We calculate the time-dependent probability density function (PDF) of the location of \emph{one} biomarker released by a cancer cell inside a cancerous blood vessel.
Subsequently, we derive the steady-state spatial concentration of the biomarkers inside the cancerous blood vessel, due to a continuous release of biomarkers. Using particle-based simulations, we show that for blood vessels of small height, such as capillaries, arterioles, and venules, the concentration of biomarkers inside the blood vessel is approximately uniform across the cross-section.
\item[] $\bullet$ Based on the results obtained for one cancerous blood vessel, we extend our model to networks comprising several blood vessels, which are part of the entire CS. To this end, we decompose the networks into three main building blocks, namely \emph{straight edges}, \emph{junction nodes}, and \emph{bifurcation nodes}, and analytically characterize the impact of each block on the steady-state concentration of the biomarkers in closed form. 
Since the numerical evaluation of the expression for the biomarker concentration obtained from our analysis may be challenging\footnote{This is due to the fact that, for computation of the biomarker concentration, the inverse of a possibly ill-conditioned matrix is needed.}, we also propose an approximation to obtain the concentration with much less complexity. In Section VI, via particle-based simulations, we show that for typical system parameter values for the blood fluid velocity and typical MNS and biomarker diffusion coefficients, the results obtained based on the approximation are accurate.
\item[] $\bullet$ We also derive the statistics of the activation levels of the MNSs traveling through the blood vessels of a given network. Furthermore, based on the proposed system model, we formulate a hypothesis testing framework to decide at the FC whether cancer cells are present in the CS or not. To this end, we derive the optimal Neyman-Pearson decision rule \cite{Poor_Book_Detection} for a given number of MNSs observed and read out at the FC by assuming that the FC knows the structure of the entire network. We then derive a simple and practical detector, which we refer to as \emph{sum detector}, that adds up the activation levels of the MNSs observed at the FC. Next, we evaluate the performance of the proposed detectors in terms of the false alarm and missed detection probabilities via Monte Carlo simulations. 
For a sample network consisting of several blood vessels, we show that while the LRT detector achieves a higher performance than the sum detector at the expense of a higher complexity, both detectors outperform a benchmark scheme with fixed nanosensors.
\end{enumerate}

The remainder of this paper is organized as follows. In Section~II, we provide the system model. In Section~III, we develop a model for cancerous blood vessels and derive the steady-state distribution of the biomarkers inside the blood vessels of a sample network of the CS. In Section~IV, we derive the statistics of the activation levels of the MNSs that pass through the network. The optimal and suboptimal detector design for the FC is presented in Section~V. Section~VI provides extensive particle-based simulation results to verify all assumptions made in the preceding sections as well as numerical results to assess the performance of the proposed detectors. Finally, Section~VII concludes the paper.

\section{System Model}\label{sysmod}
In this section, we introduce the system model considered in this paper.
We consider a part of the CS comprising several blood vessels. The blood vessels may be interpreted as the edges of a network. The edges may merge, thus forming nodes, cf. Fig. \ref{Fig.3}.
We assume that in one specific point of the network, proteins are released due to the presence of cancer cells. These proteins serve as biomarkers for the cancer. Upon release, the biomarkers move through the network driven by blood flow and diffusion.
%
%
We inject $M_{\text{max}}$ MNSs at one specific site into the CS. Each MNS can be potentially activated to a varying degree by biomarkers. In addition, we assume an FC capable of reading the MNSs' states when they reach the FC. Based on FC observations, we construct a hypothesis testing framework
and denote $\HH$ and $\HHH$ as the hypotheses for the normal (absence of cancer cells) and abnormal (presence of cancer cells) status, respectively. 
Then, the aim is to reliably decide at the FC whether $\HH$ or $\HHH$ is true based on the observations read out from the MNSs. 

In the following, we develop a simplified model for the CS suitable for formulating a decision problem. Then, we provide a model for biomarker detection at the MNSs.

\subsection{A Simplified Model for the Cardiovascular System}\label{sec:BCS}
\vspace*{-0.5mm}
In general, the CS is highly complex. Therefore, in order to keep our analysis tractable, we are interested in developing a simplified model, which retains the relevant characteristics of the CS. To this end, we consider a similar model as the one proposed in \cite[Ch. 20]{Martini}, where a particle released into the CS can enter a limited number of main paths and move through them.
Furthermore, we assume that not only cancerous cells release biomarkers but also healthy cells \cite{Hori}. The biomarkers secreted by healthy cells can be interpreted as environmental noise. 
In addition, instead of considering a complex 3-D model for the CS, to facilitate first-order insights about the propagation of biomarkers and MNSs through the CS, we consider a simplified 2-D model for the network, as depicted in Fig. \ref{Fig.3}. 
The low-dimensional model for the CS makes it not only possible to gain general physical insight into the dynamics of the CS \cite{Malatos}, but also facilitates the derivation of closed-form expressions for the biomarker concentration and the activation levels of the MNSs for the proposed MNS-based approach for the anomaly detection problem. 
\subsection{Cancerous Blood Vessel}\label{sec:blo}
\begin{figure}[!t]
 \centering
 \includegraphics[scale=0.43]{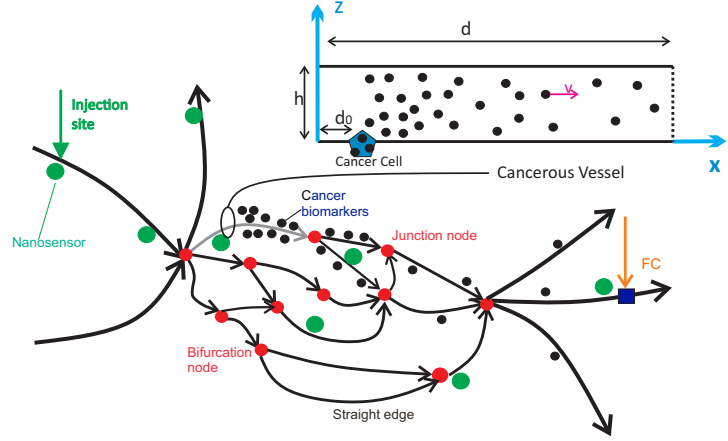}\vspace{-0.5cm}
 \caption{A schematic presentation of a sample 2-D network.}\vspace{-0.5cm}
\label{Fig.3}
\end{figure}
%
We adopt a simple model for the cancerous blood vessel. In particular, for the blood vessels, we assume a 2-D rectangular shape with length $d$ and height $h$ and reflective boundaries. The cancer biomarkers are released into one of the blood vessels at position $(x,z)=(d_0, 0)$, as depicted in Fig. \ref{Fig.3}, and undergo advection in $x$-direction and independent diffusion in $x$- and $z$-directions.
We note that the blood in small vessels like capillaries can be modeled as a non-Newtonian Casson fluid \cite{Venka}, which imposes a non-uniform flow velocity along the $z$-axis. However, for tractability of the analysis, and similar to \cite{Nakano_Let,Gentile}, and \cite{Gill}, we assume a uniform flow with constant velocity $\mathsf{v}$. 
Furthermore, we assume that the cancer resides in one specific capillary while the released biomarkers spread into all neighboring edges and nodes. We also assume that the biomarkers degrade at a rate of $\K$ [s$^{-1}$] and are released at a constant rate of $\mu$ [biomarkers$\cdot$s$^{-1}$] into the blood vessel \cite{Chen, Hori}. The release rate depends on the stage of the cancer, i.e., in the early stages, it is very low, and in the final stages, it can be very high.
For the flow velocity change at the interconnections of the vessels, i.e., the nodes of the network, we also assume uniform flow, where the velocity of the incoming and outgoing flows abruptly changes \cite{Nakano_Let}, due to the conservation of mass. The mathematical model of the biomarker release rate and the biomarker distribution inside the CS will be presented in Section \ref{sec:Bio}.

\subsection{Mobile Nanosensors}\label{SM}
MNSs are employed to decide on the presence of cancer cells in a network of interconnected blood vessels.
We assume that the MNSs are injected at one specific site of the network and that each MNS has a small area, $S$, and can measure the number of biomarkers within this area \cite{Chen}. In addition, we assume that the MNSs are passive with respect to the biomarkers, i.e., the MNSs and biomarkers move independently in the environment and do not interact with each other. In particular, an MNS can be a natural or an engineered cell which can periodically and locally measure and count the number of biomarkers inside its surface \cite{Nakano, Chen, Atkinson, Kramer, Danino, Endres, Rolfe}. 
The number of biomarkers inside an MNS can be determined either directly by counting the number of biomarkers bound by suitable cell-surface receptors \cite{Endres}, or indirectly by measuring other parameters, such as the pressure and temperature of the medium in the vicinity of an MNS's surface \cite{Rolfe}.

After MNS injection, based on the direction of the flow in the blood vessel network and the topology of the network, the MNSs follow different routes where they may eventually encounter the cancer biomarkers, e.g., when passing through cancerous tissue and capillaries. The MNSs enter the cancerous vessel at positions $(x,z)=(0,z_0)$, where we model $z_0$ as random and uniformly distributed in $[0,h]$. 
As mentioned above, we assume that each MNS can measure the number of biomarkers with a given periodicity. Similar to \cite{Adam}, we assume the time interval between two consecutive measurements is large enough such that the observations are independent. 
Furthermore, we assume that the MNS can accumulate successive observations and the FC can read out the summation value. We will refer to this value as the \emph{activation level} and denote it for the $m$-th MNS observed at the FC by random variable (RV) $A_m$ and its realization by $a_m$, respectively. The probability mass function (PMF) of this RV will be mathematically derived in Section \ref{sec:Act}.

\subsection{Fusion Center}\label{FC}
The MNSs can pass through different routes of the network and some may not arrive at the FC. Since for a given MNS the selection of the path through the network is random, a released MNS is observed at the FC with a certain probability $\rho$. Intuitively, $\rho$ can be increased by choosing a favorable location for the FC, increasing the observation window, or improving the reception process at the FC. This can be achieved, for example, by employing magnetic MNSs, which can be efficiently attracted to the FC by holding them in place by an external magnet, see \cite{Cohen} and \cite{Xie}. 
Here, we assume that the FC can observe and read the activation levels of all MNSs that come in contact with it. 
After reading out the MNSs' measurements, the FC makes a decision regarding the presence of cancer cells. Optimal and suboptimal decision rules for the FC will be provided in Section \ref{Statis}. 
%
%
%
%
%
%
%
\section{Biomarker Distribution}\label{sec:Bio}
In this section, first we consider a single cancerous blood vessel and derive the distribution of the biomarkers in it as a function of time and space. Then, we extend the obtained results to a sample network of the CS and analyze the distribution of the biomarkers in the interconnected blood vessels. 

\subsection{Biomarker Production Rate}
In the literature, several models for the production rate of cancer cells inside a blood vessel have been proposed \cite{Norton}. However, most of these models do not include the production rate of biomarkers secreted by cancerous and healthy cells. One recently established model for the production rates of cancerous and healthy cells as well as the biomarkers secreted by these cells is given in \cite{Hori}. This model can be summarized as follows.
\begin{enumerate}

\item The total number of cancer cells inside the cancerous tissue, denoted by $N_{\text{C}}(t)$, follows a Gompertzian function\cite{Norton}
\begin{align}\label{eq1}
N_{\text{C}}(t) = N_{\text{C},0}~\Ex \left( \frac{k_{\text{Gr}}}{k_{\text{Dec}}} \left( 1- \Ex\left( -k_{\text{Dec}}t \right) \right) \right),
\end{align}
where $N_{\text{C},0}$ is the initial number of cancer cells at time zero, and $k_{\text{Gr}}$ and $k_{\text{Dec}}$ are the fractional growth and decay rates of the cancer cells, respectively. For $k_{\text{Dec}}t \ll 1$, i.e., in the early stage of a cancer, (\ref{eq1}) can be approximated by
\begin{align}\label{number_cancer}
N_{\text{C}}(t) = N_{\text{C},0}~\Ex \left( k_{\text{Gr}} t \right).
\end{align}
\item Unlike the cancer cells, the number of healthy cells is assumed to be constant over time, i.e., $N_{\text{H}}(t)= N_{\text{H},0}$.

\item Cancerous cells and healthy cells secrete the same biomarkers into the medium with constant shedding rates $R_{\text{C}}$ and $R_{\text{H}}$, respectively, where $R_{\text{C}} \gg R_{\text{H}}$. However, in the vasculature, on average fractions of $f_{\text{H}}$ and $f_{\text{C}}$ of the biomarkers secreted by the healthy and cancerous cells, respectively, will be present and the remaining fractions of biomarkers will be secreted into tissue. 

\item The biomarkers degrade uniformly in the blood vessel over time with constant degradation rate $\K$. 

\end{enumerate}
Based on the above assumptions, the release rate of the biomarkers secreted by healthy cells is constant and equal to $f_{\text{C}}R_{\text{C}} N_{\text{H},0}$. However, for the cancer cells, we obtain the rate of biomarker release, denoted by $\mu$ [biomarkers$\cdot$s$^{-1}$], at time $T$ as follows
\begin{align} \label{mu}
\mu(T) \triangleq f_{\text{C}}R_{\text{C}} N_{\text{C},0}~\Ex \left( k_{\text{Gr}} T \right).
\end{align}

For healthy cells, which cover the entire surface of the blood vessels, it can be shown that the total number of secreted biomarkers that are present in the environment reaches a steady-state value of $f_{\text{H}}R_{\text{H}} N_{\text{H},0}/\K$ \cite{Hori}. 
Moreover, at any location of the CS, we can model the number of the biomarkers within a small area $S$ by a Poisson PMF \cite{Yilmaz_Poiss,Reza_Receiver}, with a mean proportional to $Sf_{\text{H}}R_{\text{H}} N_{\text{H},0}/\K$. We denote this mean by $S\xi$, which does not depend on $x$ nor $z$. 
On the other hand, for cancer cells, the number of secreted biomarkers does not reach a steady state and increases over time. Nevertheless, since the production rate $k_{\text{Gr}}$ is very low ($k_{\text{Gr}}<10^{-6}~$s$^{-1}$\cite{Brown}), using (\ref{number_cancer}) we can model the number of cancer cells during a short time interval around $T$ (e.g. a few days), denoted by $T_{\text{steady}}$, as constant. This implies a quasi-steady state where the secretion rate is approximately constant and equal to $\mu(T)$ during time interval $[T-T_{\text{steady}}/2,~T+T_{\text{steady}}/2]$. 
%
%
In the following subsections, we derive the spatial and temporal distribution of the biomarkers under the quasi-steady state assumption with constant release rate (\ref{mu}).

\subsection{Local Cancerous Blood Vessel}\label{Sense_Prob}
In this subsection, we consider a single cancerous blood vessel and evaluate the distribution of the biomarkers secreted by the cancer cells inside the blood vessel as a function of time and space. This analysis is needed to determine the PMF of the activation level of an MNS that enters the cancerous blood vessel. 

In small blood vessels, the blood in the $x$-direction can be modeled as a non-Newtonian Casson fluid, which imposes a non-uniform flow velocity profile along the $z$-axis \cite{Venka}, denoted by $v(z)$. Therefore, considering that the biomarkers are released by the cancer cells at position $(d_0,0)$ into the medium with a release rate of $\mathsf{s}(t;x,z)=\mu\delta(x-d_0)\delta(z)$, where $\delta(\cdot)$ is the delta function, the following advection-diffusion equation describes the 2-D biomarker concentration 
\begin{align}\label{diffusion}
\frac{\partial \CT(t;x,z)}{\partial t} &= \Db \nabla ^2 \CT(t;x,z) - \nabla\cdot \left( v(z) \CT(t;x,z) \right) - \K\CT(t;x,z) + \mathsf{s}(t;x,z),
\end{align}
where $\nabla\cdot (\cdot)$ is the divergence operator, $\nabla^2$ is the Laplace operator, and $\Db$ is the diffusion coefficient of the biomarkers.
Since $v(z)$ is a function of $z$, usually numerical methods are used to solve (\ref{diffusion}) \cite{Siegel}. However, as mentioned in Section \ref{sysmod}, in this paper, to gain first-order insight, we consider a 2-D model with uniform flow, i.e., $v(z)=\mathsf{v}$, which allows us to analytically solve the differential equation in (\ref{diffusion}). In particular, for uniform flow, the advection-diffusion equation for the concentration of the biomarkers inside the blood vessel simplifies to 
\begin{align}\label{diff_eq_new}
\frac{\partial \CT(t;x,z)}{\partial t} &= \Db \nabla ^2 \CT(t;x,z) - \mathsf{v} \nabla\cdot \left( \CT(t;x,z) \right) - \K\CT(t;x,z) + \mathsf{s}(t;x,z),
\end{align}
with boundary condition
\begin{align}\label{boundary}
\frac{\partial \CT(t;x,z)}{\partial z} = 0, ~~\qquad \text{for}~~ z=0,h.
\end{align}
To solve (\ref{diff_eq_new}), we first derive the spatial PDF for impulsive release of one biomarker at time $t=0$, and then use it to determine the spatial distribution of the biomarkers when they are released with constant rate $\mu$. To this end, we consider one biomarker release at time $t=0$, i.e., $\mathsf{s}(t;x,z) = \delta(t)\delta(x-d_0)\delta(z)$ and denote the solution of (\ref{diff_eq_new}) for this initial condition and the boundary condition in (\ref{boundary}) by $\mathsf{p}(t;x,z)\mathsf{u}(t)$. This solution is the spatial PDF of the released biomarker over time, where $\mathsf{u}(t)$ is the unit step function. 
Since for the considered model, the biomarkers' movements in the $x$- and $z$-directions are independent, we obtain \cite{Wayan}
\begin{align} \label{joint}
\mathsf{p}(t;x,z) &=\underbrace{
\frac{1}{\sqrt{4\pi \Db t } } \Ex \left( -\frac{\left(x-d_0-\mathsf{v}t\right)^2}{4\Db t} \right) 
}_{x\text{-direction}} \nonumber \\
&\times \underbrace{ \left( \frac{1}{h}+ \frac{2}{h} \sum_{k=1}^{\infty} \Ex\left( -\Db \left( \frac{k\pi}{h} \right)^2 t \right) \mathsf{cos} \left( \frac{k\pi z}{h} \right) 
\right) }_{z\text{-direction}} \underbrace{\Ex \left( - {\K}t \right)}_{\text{degradation}},
\end{align}
where $h$ is the height of the cancerous blood vessel. Then, the concentration of the biomarkers in the blood vessel at time $t$ and location $(x,z)$ can be obtained as follows
\begin{align}\label{mean}
\CT(t;x,z) = \int_{0}^{t} \mu\hspace{0.07cm} \mathsf{p}(t-\tau;x,z)\mathsf{u}(t-\tau)\mathsf{d}\tau.
\end{align}
Given (\ref{joint}), in the following lemma, by evaluating (\ref{mean}) we provide a closed-form expression for the steady-state concentration of the biomarkers secreted by the cancer cells.
\begin{lemma}\label{Lemma1}
For a single cancerous vessel with cancer cells located at $(d_0,0)$ and biomarker secretion rate $\mu$, the steady-state concentration of cancer biomarkers at $(x,z)$ is given by
\begin{align}\label{final_mean}
\CT(x,z) &= \int_{t=0}^{\infty} \mu\hspace{0.07cm} \mathsf{p}(t;x,z)\mathsf{d}t \nonumber \\
&= \mu~ \frac{\Ex \left( \frac{(x-d_0)\mathsf{v}}{2\Db} \left( 1- \mathsf{sgn}(x-d_0) \sqrt{1+\frac{4\Db \K}{\mathsf{v}^2}} \right) \right)}{h \mathsf{v} \sqrt{1+\frac{4\Db \K}{\mathsf{v}^2}} } \nonumber \\
&+ 2\mu \sum_{k=1}^{\infty} \frac{\mathsf{cos}\frac{k\pi z}{h} ~\Ex \left( \frac{(x-d_0)\mathsf{v}}{2\Db} \left( 1- \mathsf{sgn}(x-d_0) \sqrt{1+\frac{4\Db \K}{\mathsf{v}^2} + \frac{4k^2\pi^2\Db^2 }{h^2\mathsf{v}^2} } \right) \right) }{h \mathsf{v} \sqrt{1+\frac{4\Db \K}{\mathsf{v}^2}+ \frac{4k^2\pi^2 \Db }{h^2\mathsf{v}^2} }},
\end{align}
where $\mathsf{sgn}(x)$ is the sign function which is $1$ for $x>0$, $0$ for $x=0$, and $-1$ for $x<0$. 
\end{lemma}
\begin{IEEEproof}
$\CT(x,z)$ is obtained by substituting (\ref{joint}) into (\ref{mean}) and taking the limit $t \rightarrow \infty$, where $\mu$ is given by (\ref{mu}). 
\end{IEEEproof}
\begin{remark}\label{rem1}
It can be shown that for small blood vessels (such as capillaries, arterioles, and venules, where $h<100~\mu$m \cite{Mary}), the steady-state concentration of the biomarkers secreted by cancer cells $\CT(x,z)$ is approximately constant with respect to $z$ for a given $x\neq d_0$. Therefore, we can approximate the concentration as
\begin{align}\label{cx_last}
\CT(x,z) = \frac{\CT(x)}{h},~0\leq z\leq h,
\end{align}
where
\begin{align}\label{cx}
\CT(x) = \int_{z=0}^{h} \CT(x,z) \mathsf{d}z = \mu~ \frac{\Ex \left( \frac{(x-d_0)\mathsf{v}}{2\Db} \left( 1- \mathsf{sgn}(x-d_0) \sqrt{1+\frac{4\Db \K}{\mathsf{v}^2}} \right) \right)}{h \mathsf{v} \sqrt{1+\frac{4\Db \K}{\mathsf{v}^2}} }.
\end{align}
This approximation is further investigated in Section~\ref{Numeric}. We use this result to solve a 1-D advection-diffusion equation in Section~\ref{Network} for a network of connected blood vessels.
\end{remark}

\begin{remark}\label{remark3}
For typical system parameters for blood vessels and biomarkers, we have ${4\Db \K}/{\mathsf{v}^2}$ $\ll1$. Hence, (\ref{cx}) can be simplified as
\begin{align}\label{velocity_change}
\CT(x) \approx \left\{ 
\begin{array}{l l}
\frac{\mu}{\mathsf{v}} \hspace*{0.5mm} \Ex \left( \frac{(x-d_0) \mathsf{v}}{\Db} \right) ~~~ &\text{if}~x \leq d_0,\\
\frac{\mu}{\mathsf{v}} \hspace*{0.5mm} \Ex \left( \frac{-(x-d_0) k_{\text{Dg}}}{\mathsf{v}} \right) ~~~ &\text{if}~x \geq d_0. \\
\end{array} \right.
\end{align}
\end{remark}

\subsection{Network of Blood Vessels}\label{Network}

In this subsection, we derive closed-form expressions for the distribution of the concentration of the biomarkers inside a given network of blood vessels. The secreted biomarkers spread across multiple blood vessels which imposes further boundary conditions on the advection-diffusion equation in (\ref{diff_eq_new}). Therefore, we cannot directly apply the results derived in (\ref{final_mean}) for one vessel to a network of several connected vessels. 
Since the dependence of the concentration of the biomarkers on $z$ is negligible when $h$ is not very large, see Remark \ref{rem1}, we approximated the distribution of the biomarker concentration along the $z$-axis as uniform (\ref{cx_last}). 
Therefore, in this subsection, we consider a 1-D blood vessel network model and analyze how junction and bifurcation nodes influence the distribution of the biomarkers in the connected blood vessels. Then, considering the uniform distribution of the biomarkers with respect to $z$, we can obtain the 2-D concentration of the biomarkers based on the 1-D concentration.

As can be observed from Fig. \ref{Fig.3}, a network typically comprises three fundamental blocks: 
\begin{enumerate}
\item[I:] \emph{Straight edges}: The blood vessels that connect two adjacent nodes.
\item[II:] \emph{Junction nodes}: In a junction node, two or more incoming blood flows join to become one single flow.
\item[III:] \emph{Bifurcation nodes}: In a bifurcation node, an incoming blood flow is split into two or more outgoing flows.
\end{enumerate}

In Section \ref{Sense_Prob}, we provided the steady-state concentration of the biomarkers for a local straight edge in (\ref{cx_last}). In the following, we study the effect of junction and bifurcation nodes on the concentration of the biomarkers inside the edges.

\subsubsection{Junction Node}\label{subJun}

\begin{figure}[!t]
 \centering
 \includegraphics[scale=0.53]{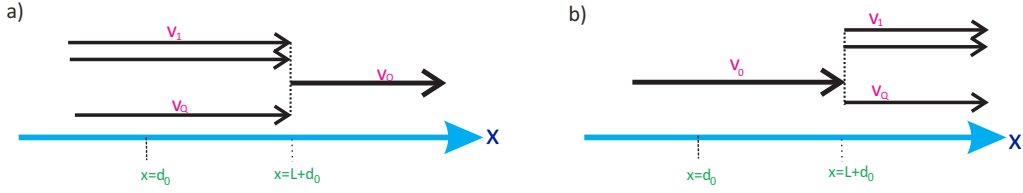}\vspace{-0.5cm}
 \caption{a) Schematic model of a junction node, b) schematic model of a bifurcation node. }\vspace{-5.0 mm}
\label{Fig.NetwNew}
\end{figure}

For a junction node, we consider the case where $Q$ small edges are joined to form a bigger edge, see Fig. \ref{Fig.NetwNew}a). We assume that the $j$-th incoming edge has flow velocity $\mathsf{v}_j$ and cross-sectional area $S_j$, and the outgoing edge has flow velocity $\mathsf{v}_O$ and cross-sectional area $S_O$. Then, based on the first law of dynamic flow, the incoming and outgoing flow rates must be identical, which results in $\mathsf{v}_O = \frac{\sum_{j=1}^{Q} S_j \mathsf{v}_j}{S_O}$\cite{Mechanics}.
Without loss of generality, we assume that the first edge with velocity $\mathsf{v}_1$ secretes cancer biomarkers at location $x=d_0$ into the medium, and is joined with the $Q-1$ other edges at $x=d_0+L$ to yield a larger edge with flow velocity $\mathsf{v}_O$. We are particularly interested in finding the concentration of biomarkers inside the first small edge and the outgoing edge. 
Since we have assumed steady-state conditions, the following differential equation describes the 1-D concentration of the biomarkers inside the first small edge and the outgoing edge 
\begin{align}
\Db \frac{\partial^2 \CT(x)}{\partial x^2} - \mathsf{v}_g \frac{\partial \CT(x)}{\partial x} - \K\CT(x) = 0, ~g=1,O, \label{eq:23} 
\end{align}
where steady-state operation was assumed, i.e., ${\partial \CT(x)}/{\partial t} = 0$.
Since $\Db$, $\mathsf{v}$, and $\K$ are all positive, the solution of (\ref{eq:23}) will be a combination of exponential functions of the form $\Ex(\lambda x)$ where the coefficients $\lambda$ must be roots of following algebraic equation \cite{Stewart}
\begin{align}
\Db \lambda^2 - \mathsf{v}_g \lambda - \K = 0, ~g=1,O,
\end{align}
which has two possible solutions.
Furthermore, based on the fundamental laws of diffusion and conservation of mass, the following two conditions have to be satisfied: i) the concentration of the molecules must be continuous, ii) the flux must be continuous and can be obtained as \cite{Roisin}
\begin{align}
\F(x) = -\Db \frac{\partial \CT(x)}{\partial x}+ \mathsf{v}_g \CT(x), ~g=1, O.
\end{align}
In the following, we analytically derive the concentration of the biomarkers and provide a suitable approximation for simpler evaluation.

\textbf{Exact Solution:}
Based on the above discussion, the steady-state concentration of the biomarkers can be shown to have the following general form \cite{Stewart}
\begin{align}\label{eq:my16}
\CT(x)=\left\{ 
\begin{array}{l l l}
B_1~ \Ex \left( \lambda_{+,1} (x-d_0) \right) ~\qquad \qquad \qquad \qquad &\text{if}~x \leq d_0, \\
B_2~ \Ex \left( \lambda_{+,1} (x-d_0) \right) + B_3~ \Ex \left( \lambda_{-,1} (x-d_0) \right) ~~~&\text{if}~ d_0 \leq x \leq d_0+L, \\
B_4~ \Ex \left( \lambda_{-,O} \left(x-d_0-L\right) \right) ~\qquad \qquad \qquad \qquad &\text{if}~ x \geq L+d_0, \\
\end{array} \right.
\end{align}
where
\begin{align}\label{junkgensol}
\lambda_{\pm,1} = \frac{\mathsf{v}_1\left( 1\pm\sqrt{1+\frac{4\Db \K}{\mathsf{v}_1^2} } \right) }{2\Db},
\lambda_{-,O} = \frac{\mathsf{v}_O\left( 1-\sqrt{1+\frac{4\Db \K}{\mathsf{v}_O^2} } \right) }{2\Db}.
\end{align}
Here, $B_1$, $B_2$, $B_3$, and $B_4$ are unknown parameters that have to be determined.
Combining (\ref{eq:23})--(\ref{junkgensol}), we arrive at the linear equation $\mathsf{M} \vec{x} = \vec{y}$ with
\begin{align}\label{main_join}
\begin{array} {c c}
\hspace*{-4.0mm} \mathsf{M} \hspace{-0.1cm}= \hspace{-0.1cm}
\begin{bmatrix} 
1 & -1 & -1 & 0 \\
\Db\lambda_{+,1} & -\Db\lambda_{+,1} & -\Db\lambda_{-,1} & 0 \\
0 & -\Ex \left( \lambda_{+,1} L \right) & -\Ex \left( \lambda_{-,1} L \right) & 1 \\
0 & -\Db \lambda_{+,1} \Ex \left( \lambda_{+,1} L \right) & -\Db \lambda_{-,1} \Ex \left( \lambda_{-,1} L \right) & \mathsf{v}_1-\mathsf{v}_O+\Db\lambda_{-,O}
\end{bmatrix}\hspace{-0.1cm}, 
\vec{x} \hspace{-0.1cm} = \hspace{-0.1cm}
\begin{bmatrix} 
B_1 \\
B_2 \\
B_3 \\
B_4
\end{bmatrix}\hspace{-0.1cm}, 
\vec{y} \hspace{-0.1cm}= \hspace{-0.1cm} 
\begin{bmatrix} 
0 \\
\mu \\
0 \\
0
\end{bmatrix} \hspace{-0.1cm}.\\
\end{array}
\end{align}
The unique solution for $\vec{x}$ can be obtained as
$\vec{x} = \mathsf{M}^{-1} \vec{y}$, where $\mathsf{M}^{-1} $ is the inverse of matrix $\mathsf{M}$.

\textbf{Approximate Solution:}
Due to the possibly large values for $\lambda_{\pm,1}$ and $\lambda_{-,O}$, matrix $\mathsf{M}$ may be ill-conditioned, which makes the determination of unknown parameter vector $\vec{x}$ numerically challenging. Hence, in the following, we propose a simple approximate solution to (\ref{eq:23}). In particular, we consider the following approximations:
\begin{enumerate}
\item[I:] Suppose that the flow in edge $1$ with velocity $\mathsf{v}_1$ is not influenced by the other connected blood vessels at $x=d_0+L$. Then, we use the solution for a local straight edge in (\ref{cx}) for $x \leq d_0+L$.
\item[II:] Consider the concentration at $x=d_0+L$ as a virtual source for the blood vessel with outgoing flow with velocity $\mathsf{v}_O$. Then, we employ (\ref{cx}) for $x\geq d_0+L$ and flow velocity $\mathsf{v}_O$.
\end{enumerate} 
Considering these approximations, the approximated solution can be obtained as follows
\begin{align}\label{appnew123}
\CT(x)=\left\{ 
\begin{array}{l l l}
\mu~ \frac{\Ex \left( \frac{(x-d_0)\mathsf{v}_1}{2\Db} \left( 1- \mathsf{sgn}(x-d_0) \sqrt{1+\frac{4\Db \K}{\mathsf{v}_1^2}} \right) \right)}{\mathsf{v}_1 \sqrt{1+\frac{4\Db \K}{\mathsf{v}_1^2}} }~~&\text{if}~ x \leq d_0+L, \\
\left( \mu ~{\Ex \left( \frac{L\mathsf{v}_1}{2\Db} \left( 1- \sqrt{1+\frac{4\Db \K}{\mathsf{v}_1^2}} \right) \right)} \right) \times 
\frac{\Ex \left( \frac{(x-d_0-L)\mathsf{v}_O}{2\Db} \left( 1- \sqrt{1+\frac{4\Db \K}{\mathsf{v}_O^2}} \right) \right)}{\mathsf{v}_O \sqrt{1+\frac{4\Db \K}{\mathsf{v}_O^2}} }
~~&\text{if}~ x \geq d_0+L. \\
\end{array} \right.
\end{align}
\begin{remark}
Similar to Remark \ref{remark3}, for typical parameter values for biomarkers and blood vessels, we can assume that ${4\Db \K}/{\mathsf{v}_1^2}\ll 1$ and ${4\Db \K}/{\mathsf{v}_O^2}\ll 1$. Hence, (\ref{appnew123}) can be simplified as
\begin{align}\label{appnew__} %
\CT(x) \approx \left\{ 
\begin{array}{l l l}
\frac{\mu}{\mathsf{v}_1} \Ex \left( \frac{(x-d_0) \mathsf{v}_1}{\Db} \right) ~~~~~ &\text{if}~~~x \leq d_0,\\
\frac{\mu}{\mathsf{v}_1} \Ex \left( \frac{-(x-d_0) \K}{\mathsf{v}_1} \right) ~~~~~ &\text{if}~~~ d_0\leq x \leq d_0+L, \\
\frac{\mu}{\mathsf{v}_O} \Ex \left( -\K \times \frac{(\mathsf{v}_O-\mathsf{v}_1)L +(x-d_0)\mathsf{v}_1}{\mathsf{v}_1\mathsf{v}_O} \right) ~~~~~ &\text{if}~~~ x \geq d_0+L. \\
\end{array} \right.
\end{align}
\end{remark}

\subsubsection{Bifurcation Node}\label{subBif} 
For a bifurcation node, we assume that one edge is divided into $Q$ edges. Furthermore, let us assume that the incoming flow velocity is $\mathsf{v}_0$ and the outgoing flow velocity for the $j$-th edge ($j=1,\ldots, Q$) is $\mathsf{v}_j$, see Fig. \ref{Fig.NetwNew}b). 
The derivation of an expression for the velocity of the outgoing flows involves both the first and second law of fluid mechanics, which, in turn, requires knowledge of the entire network topology \cite{Mechanics}. Therefore, similar to Section \ref{subJun}, we assume that the flow velocities of the edges are given. Then, using the principle of mass conservation \cite{Mechanics}, we can derive the probability that a biomarker secreted at $x=d_0$ in the incoming edge, enters the $j$-th outgoing edge at $x=d_0+L$ as follows 
\begin{align}\label{vel_bif}
p_j = \frac{S_j \mathsf{v}_j}{\sum_{j=1}^{Q}S_j \mathsf{v}_j},~j \in \{1, \cdots, Q \},
\end{align} 
where $S_j$ is the cross-sectional area of the $j$-th outgoing edge.
By using a similar approach as in Section \ref{subJun}, we arrive at the following analytical expression for the concentration of the biomarkers at the bifurcation node
\begin{align}\label{bifu} %
\CT(x)=\left\{ 
\begin{array}{l l l}
\hspace*{-2.0mm}C_1~ \Ex \left( \lambda_{+,0} (x-d_0) \right)   &\text{if}~x \leq d_0, \\
\hspace*{-2.0mm}C_2~ \Ex \left( \lambda_{+,0} (x-d_0) \right) + C_3~ \Ex \left( \lambda_{-,0} (x-d_0) \right) &\text{if}~ d_0 \leq x \leq d_0+L, \\
\hspace*{-2.0mm}p_jC_4~ \Ex \left( \lambda_{-,j} \left(x-d_0-L\right) \right)   &\text{if}~ x \geq d_0+L,~ \forall j\in \{1, \cdots, Q\}, \\
\end{array} \right.
\end{align}
where $\lambda_{+,0}$ and $\lambda_{-,0}$ are the positive and negative roots of
$\Db \lambda^2 - \mathsf{v}_0\lambda - \K = 0$, respectively, and $\lambda_{-,j}$ is the negative root of $\Db \lambda^2 - \mathsf{v}_j\lambda - \K = 0$. 
Similar to the approach proposed in Section \ref{subJun}, by considering the initial value and boundary conditions at $x=d_0$ and $x=d_0+L$, we can construct a linear system of equations to determine $C_1, C_2, C_3$, and $C_4$ as well as the exact and approximate solutions for $\CT(x)$ in \eqref{bifu}, which we omit here due to space limitation. 

Based on the results obtained in this section, for any given network, inside each blood vessel of the network, we can derive the concentrations of the biomarkers, $\CT(x,z)$, secreted by cancer cells. That is, by plugging the 1-D expressions into (\ref{cx_last}) we obtain the 2-D concentration of the biomarkers.

\subsection{PMF of the Number of Biomarkers}
In the following, we derive the spatial PMF of the number of biomarkers inside a small virtual observation area $S$ representing one MNS. In the next section, we use this PMF to derive the activation level of the MNSs. 
With $\CT(x)$ derived in Sections \ref{Sense_Prob} and \ref{Network} and assuming that $S$ is small so that the concentration of biomarkers is approximately uniform inside the observation area, the mean number of biomarkers inside the area $S$ centered at position $(x,z)$ is given by $S\CT(x)/h$. However, this does not directly give insight about the PMF of the random number of biomarkers inside the observation area.
In the following, we show that the number of biomarkers inside this area generally follows a \emph{Poisson Binomial} distribution. 

Let us denote the PMF for having at time $t$, $n$ biomarkers out of the $N=\mu t$ biomarkers released in total, which undergo advection and diffusion and may degrade until time $t$, inside the observation area centered at $(x,z)$ by $\mathsf{p}_{\text{bio},n}(t;x,z)$. 
To derive $\mathsf{p}_{\text{bio},n}(t;x,z)$, suppose that the release time for the $n$-th biomarker, $n\in \{1, \ldots, N \}$, is $T_n$. Hence, the probability of having the $n$-th biomarker inside the observation area at time $t$ is given by $S\mathsf{p}(t-T_n;x,z) \mathsf{u}(t-T_n)$, where $\mathsf{p}(\cdot;\cdot,\cdot)$ is given in (\ref{joint}).
Since the biomarkers have different release times, the number of biomarkers inside the observation area at time $t$ is a Bernoulli trial with different success probabilities for different biomarkers, which leads to a Poisson Binomial distribution.
Hence, the probability of simultaneously having $n$ of the released biomarkers inside the observation area can be expressed as
\begin{align}\label{eq:17}
\mathsf{p}_{\text{bio},n}(t;x,z) = \sum_{{B}\in \mathsf{F}_n} \prod_{i \in {B}}S\mathsf{p}(t-T_i;x,z) \prod_{i \in {B}^c} \left( 1- S\mathsf{p}(t-T_i;x,z) \right),
\end{align}
where $\mathsf{F}_n$ is the set of all subsets of $n$ integers $\{1,2,3,...,N\}$, with $\big|\mathsf{F}_n\big| = {N \choose n}$, where $\big|\cdot\big|$ denotes the cardinality of a set .

The Poisson Binomial distribution is cumbersome to work with, but can often be approximated by a Poisson distribution when the number of trials is high and the success probability is small, cf. \cite{Binomial}. 
To this end and for simplicity, we approximate (\ref{eq:17}) by a Poisson PMF with mean $S\CT(x)/h$. 
For our system model, this approximation is accurate when the number of released biomarkers is high; and for each biomarker, the probability of being observed inside the MNS is small. 
In Section~\ref{Numeric}, we validate the accuracy of this approximation via particle-based simulation. Therefore, we model the PMF for the number of biomarkers, secreted by cancerous and healthy cells, found inside a small observation area $S$ centered at $(x,z)$, as
\begin{align}\label{Nbio}
N_{\text{bio}}(x) \sim \Po \left( S\frac{\CT(x)}{h}+S\xi \right),
\end{align}
where $\Po(\gamma)$ denotes a Poisson distribution with mean $\gamma$. Here, we have denoted the number of biomarkers, secreted by healthy cells, inside the observation area by a Poisson RV with mean $S\xi$.

\vspace*{-3mm}
\section{MNS Activation Level Statistic}\label{sec:Act}

In the previous section, we derived the concentration of the biomarkers inside a network of blood vessels. Using this result, we then derived the PMF of the number of biomarkers inside a small area in the network. In this section, we use the results of Section \ref{sec:Bio}, to determine the PMF of the activation levels of the MNSs, which is needed for detection at the FC.

To derive the PMF of the activation level of the MNSs, we have to consider all possible routes inside the network between the injection site and the FC as well as the distribution of the biomarkers inside each edge of the network. To formulate this rigorously, we assume that the network between the injection site and the FC contains $L$ different edges, where the $l$-th edge, $l \in \{1, \dots, L\}$, is denoted by E$l$. Then, based on the topology of the network and the flow directions inside the vessels, we assume that there exist $R$ different routes starting at the injection site and ending at the FC. We denote the set that contains the indices of the edges that construct the $r$-th route, $r\in \{1, \dots, R \}$, by $\mathcal{R}_r$. In general, which route an MNS moves through is not a priori known and is modelled as an RV. Therefore, for the $m$-th observed MNS at the FC, we model the route that it has taken by RV $G_m$, where $G_m \in \{1, 2, \ldots, R \}$. We denote the probability $\msP(G_m = r)$ that the $m$-th MNS passes through the $r$-th route by $q_r$. We also define $\CT^{\{l\}}(x)$ as the concentration of the biomarkers at local position $x$ inside the $l$-th edge, which depending on the topology of the network, can be obtained from (\ref{cx}), (\ref{appnew123}), or (\ref{bifu}). 
Furthermore, based on the fact that each MNS takes measurements periodically and the lengths of the routes, in general, are not the same, the measurement time inside each edge depends on the route $r$ that the MNS travels along. We denote the set of time instances at which an MNS traveling via the $r$-th route takes measurements inside the $l$-th edge by $\mathcal{T}^{\{l\}}_{r}$. We note that since the $l$-th edge may not be included in $\mathcal{R}_r$, set $\mathcal{T}^{\{l\}}_{r}$ can be empty. 
Furthermore, we denote the number of biomarkers that the $m$-th MNS observes inside the $l$-th edge at time $t$ by RV $N_m^{\{l\}}(t)$, which has a Poisson PMF, cf. (\ref{Nbio}).
Now, based on the definition of the activation level $a_m$ given in Section \ref{SM}, we can express the PMF for the activation level of the $m$-th MNS as 
\begin{align}\label{gen_act}
\msP ({A}_m = a_m) &= \sum_{r=1}^{R} \msP\left( G_m = r \right) \msP \left({A}_m = a_m\big| r\right) \nonumber \\
&= \sum_{r=1}^{R} \msP\left( G_m = r \right) \msP\left( \sum_{l\in \mathcal{R}_r} \sum_{t\in \mathcal{T}^{\{l\}}_{r}} N_m^{\{l\}}(t) = a_m \right) \nonumber \\
&= \sum_{r=1}^{R} q_r \msP\left( \sum_{l\in \mathcal{R}_r} \sum_{t\in \mathcal{T}^{\{l\}}_{r}} N_m^{\{l\}}(t) = a_m \right).
\end{align}
To evaluate (\ref{gen_act}), first we need to determine the PMF of $N_m^{\{l\}}(t)$. However, since the position of the MNS in $x$-direction is generally not deterministic due to diffusion, to derive the PMF of $N_m^{\{l\}}(t)$, we employ a simplification. In particular, due to the small value of the diffusion coefficient, $D_{\text{n}}$, of typical MNSs \cite{Hori}, advection is dominant over diffusion regarding the movement of the MNSs along the $x$-axis. For instance, for capillaries, normally $\mathsf{v}\geq0.03~$cm$\cdot$s$^{-1}$, $d \approx 9\times 10^{-2}~$cm, and $D_{\text{n}} \leq 10^{-8}~$cm$^2 \cdot$s$^{-1}$ holds \cite{Hori, Chen, Marieb}. 
In this case, the distance that an MNS can travel by advection during a test time interval of $T_{\text{test}} \triangleq d/v = 0.09/0.03=3~$s, is $d = 9\times 10^{-2}~$cm which is much larger than the standard deviation of the distance that the MNS typically moves by diffusion during the same time interval, $\sqrt{ 2D_{\text{n}}T_{\text{test}}}$ $= 2.4\times 10^{-4}~$cm\footnote{We note that since for biomarkers, normally $\Db \geq 10^{-6}$cm$^2$s$^{-1}$ \cite{Chen, Hori}, the standard deviation of the distance that a biomarker moves by diffusion during $T_{\text{test}}$ is larger than $2.4 \times 10^{-3}~$cm and not negligible compared to the distance that the biomarker moves by advection.}. 
Hence, the locations where measurement are taken along the $x$-axis are practically deterministic. Thus, we define the set of deterministic measurement positions for an MNS during its journey inside the $l$-th edge via the $r$-th route by $\mathcal{X}^{\{l\}}_{r}=\left\lbrace x\big| x= \mathsf{v}_lt, \forall t\in \mathcal{T}^{\{l\}}_{r} \right\rbrace$, where $\mathsf{v}_l$ is the flow velocity in the $l$-th edge. Therefore, since an MNS's measurements are assumed independent, based on the thinning property of Poisson processes, we obtain the activation level conditioned on the MNS having taken the $r$-th route as
\begin{align}\label{cond_act}
A_m \big| r = \sum_{l\in \mathcal{R}_r} \sum_{t\in \mathcal{T}^{\{l\}}_{r}} N_m^{\{l\}}(t) \sim \Po \left(\sum_{l\in \mathcal{R}_r} \sum_{x\in \mathcal{X}^{\{l\}}_{r}} \left( \underbrace{ S\frac{\CT^{\{l\}}(x)}{h}}_{\text{Cancerous cells}} + \underbrace{S\xi}_{\text{Healthy cells}} \right) \right). 
\end{align}
Note that in (\ref{cond_act}) the contributions of both cancerous and healthy cells are included.
%
%
For ease of presentation, we define $\J_r$ and $\A_r$ as the mean values of the conditional RV $A_m \big| r$ under hypotheses $\HH$ and $\HHH$, respectively, which can be written as
\begin{align}\label{mean_act_new}
\left\{ 
\begin{array}{l l}
\J_r &\triangleq S \xi \sum_{l\in \mathcal{R}_r} \big| \mathcal{X}^{\{l\}}_{r} \big|\\
\A_r &\triangleq \sum_{l\in \mathcal{R}_r} \sum_{x\in \mathcal{X}^{\{l\}}_{r}} \left( S\frac{\CT^{\{l\}}(x)}{h}+S\xi \right). \\
\end{array} \right.
\end{align}
Therefore, the activation level given in (\ref{gen_act}) can be represented for hypotheses $\HH$ and $\HHH$ as follows
\begin{align}\label{act_end}
\msP ({A}_m = a_m) = \left\{ 
\begin{array}{l l}
\sum_{r=1}^{R} q_r \frac{\Ex \left(-\J_r \right) \left(\J_r\right)^{a_m} }{a_m!},~~&\text{under}~\HH, \\
\sum_{r=1}^{R} q_r \frac{\Ex \left(-\A_r \right) \left(\A_r\right)^{a_m} }{a_m!},~~&\text{under}~\HHH. \\
\end{array} \right.
\end{align}

\section{Detector Design at FC}\label{Statis}

Based on the results derived in Section~\ref{sec:Act}, in this section, we derive the optimal decision rule at the FC with respect to the Neyman-Pearson criterion \cite{Poor_Book_Detection}, which is referred to as the \emph{likelihood ratio test} (LRT) and relies on the knowledge of the network topology between the injection site and the FC. However, in practice, we do not know the biomarker secretion rate and the location of the cancer cell. Therefore, we also propose a simple suboptimal detector that does not require knowledge of the network topology. The optimal LRT decision rule serves as an upper bound for the performance of any suboptimal detector for the proposed MNS-based system.
We note that the number of MNSs observed at the FC follows a Binomial distribution with $M_{\text{max}}$ trials and success probability $\rho$, $\mathsf{Bi}(M_{\text{max}},\rho)$, cf. Section~\ref{FC}. 
%
Here, we assume that the FC makes a decision after observing $M$ MNSs out of all $M_{\text{max}}$ injected MNSs.

Let $\vec{A}_{M}$ denote an RV vector modeling the vector containing the activation levels of all $M$ MNSs observed during the observation window, i.e., $\vec{A}_{M}=[{A}_1, \dots, {A}_M]^T$, and let $\vec{a}_{M}$ be a realization of $\vec{A}_{M}$.
According to the Neyman-Pearson criterion, the goal is to design the optimal detector that minimizes the missed detection probability subject to a pre-assigned upper bound $\alpha$ on the false alarm probability, i.e.,
\begin{align}\label{criteria}
&\min_{\text{decision rules}} \Pm ,~ \text{subject to}~ \Pf\leq \alpha,
\end{align}
where $\Pf$ and $\Pm$ are the false alarm and missed detection probabilities, respectively.
The solution to (\ref{criteria}) is the well-known LRT \cite{Poor_Book_Detection} that compares the log-likelihood ratio (LLR) with a threshold denoted by $\tau$, where $\tau$ is chosen such that $\Pf=\alpha$. 
In particular, the decision rule for the LRT can be characterized as
\begin{align} \label{mMAP_ns}
\text{d}_{\text{optimal}}=\left\{ 
\begin{array}{l l}
0, & \,\, \text{if $~ \overline{\mathsf{LLR}}(\vec{a}_{M}) \leqslant \tau$}, \\
1, & \,\, \text{otherwise}, \\
\end{array} \right.
\end{align} 
where $\text{d}_{\text{optimal}}=j$ means that the detector selects hypothesis $\HTemp_j,\,\,j=0,1$. In (\ref{mMAP_ns}), $\overline{\mathsf{LLR}}(\vec{a}_{M})$ is given by
\begin{align} \label{LLR}
\overline{\mathsf{LLR}}(\vec{a}_{\hspace{0.05cm}M}) = \mathsf{log}\left(\frac{\msP \left( \vec{A}_{M}=\vec{a}_{M} | \HHH\right)}{\msP \left( \vec{A}_{M}=\vec{a}_{M} | \HH \right)}\right).
\end{align}
In the following, we derive $\overline{\mathsf{LLR}}(\vec{a}_{M})$ for the problem at hand. Since the movements of different MNSs are independent, the activation levels of different MNSs are statistically independent, which results in
\begin{align}\label{LLR00}
\msP \left( \vec{A}_{M}=\vec{a}_{M} | \HTemp_{j}\right) = \prod_{m=1}^{M} \msP \left( A_m=a_m | \HTemp_{j}\right),~~j=0,1.
\end{align}
Therefore, based on \eqref{LLR} and \eqref{LLR00}, we can express $\overline{\mathsf{LLR}}$ as
\begin{align}\label{LLR2}
\overline{\mathsf{LLR}}(\vec{a}_{M}) = \sum_{m=1}^{M} \mathsf{log}\left(\frac{\msP \left( A_m=a_m | \HHH\right)}{\msP \left( A_m=a_m | \HH \right)}\right) \triangleq \sum_{m=1}^{M} \mathsf{LLR}(a_m), 
\end{align}
where using (\ref{act_end}) we obtain
\begin{align}\label{LLR_loc}
\mathsf{LLR}(a_m) &= \mathsf{log} \left( \frac{ \sum_{r=1}^{R} q_r~ \Ex \left( - \A_r \right) \A_r^{a_m} }{\sum_{r=1}^{R} q_r~ \Ex \left( - \J_r \right) \J_r^{a_m}} \right) 
\end{align}
for the LLR of the observation of the $m$-th MNS. 

Since in reality the network topology, and hence the values of $\A_r$ and $\J_r$ are not known at the FC, we propose a simple alternative but suboptimal detector that does not need knowledge of the network topology. The decision rule of this detector, which we refer to as \emph{sum detector}, is as follows
\begin{align} \label{practical}
\text{d}_{\text{sum detector}}=\left\{ 
\begin{array}{l l}
0, & \,\, \text{if $~ \sum_{m=1}^{M}a_m \leqslant \tau'$}, \\
1, & \,\, \text{otherwise}, \\
\end{array} \right.
\end{align} 
where $\tau'$ is chosen such that the decision rule in (\ref{practical}) yields $\Pf=\alpha$.
In Section \ref{sec:performance}, we evaluate the performance of the proposed detectors in terms of $\Pf$ and $\Pm$.


\vspace{-0.2cm}
\section{Simulation Results}\label{Numeric}

In this section, we first validate the system model and the assumptions made for the analysis presented in this paper via particle-based simulation. Then, we provide Monte Carlo simulation results to assess the performance of the proposed detection schemes.

\vspace*{-0.0mm}
\subsection{Model and Assumption Verification}
We first consider a single blood vessel which is sufficient to verify most of our assumptions. Subsequently, we introduce an example network for a more thorough analysis.

\subsubsection{Single Blood Vessel}
To verify our analysis, we perform a 2-D particle-based simulation of both the MNSs and the biomarkers secreted by the cancer cells. Table \ref{table:2} summarizes the system parameters that are used for all simulations, unless stated otherwise.
\begin{table}[!t]
\renewcommand{\arraystretch}{0.9}
\centering
\vspace{-0.6cm}
\caption{List of Default Simulation Parameters \cite{Hori}. }
\vspace{-0.4cm}
\begin{tabular}{|l | l || l | l |} 
\hline
Parameter & Value & Parameter & Value\\ [0.1ex] 
\hline
$\Db$ & $10^{-6}$cm$^2$s$^{-1}$ &$D_{\text{n}}$ & $10^{-8}$cm$^2$s$^{-1}$ \\
\hline
$k_{\text{Deg}}$ & $2\times 10^{-4}$s$^{-1}$ &$\mathsf{v}$ & $3\times 10^{-2}$cm~s$^{-1}$\\
\hline
$k_{\text{Gr}}$ & $4.63\times 10^{-7}$s$^{-1}$ & $\{\mu,\xi\}$ & $\{1~$s$^{-1}, 10^{4}~$cm$^{-2}\}$ \\
\hline
$k_{\text{Dec}}$ & $1.49\times 10^{-11}$s$^{-1}$ &$N_{\text{C},0}$ & $1$\\ 
\hline
$f_{\text{C}}R_{\text{C}}$ & $10^{-1}$ &$h$ & $4\times 10^{-3}$~cm \\ 
\hline
$S$ & $10^{-5}$~cm$^2$ &$d_0$ & $0$~cm \\ 
\hline
\end{tabular}
\label{table:2}
\vspace{-0.7cm}
\end{table}
Simulation results are averaged over $10^3$ independent realizations of the biomarker release, where for each realization, we assume that the biomarkers are released into the blood vessel with rate $\mu=1~$s$^{-1}$. This rate is calculated via (\ref{mu}) for $T\approx2$~months, i.e., the early stage of a cancer. 

\begin{figure}[!t]
 \centering
 \includegraphics[scale=0.465]{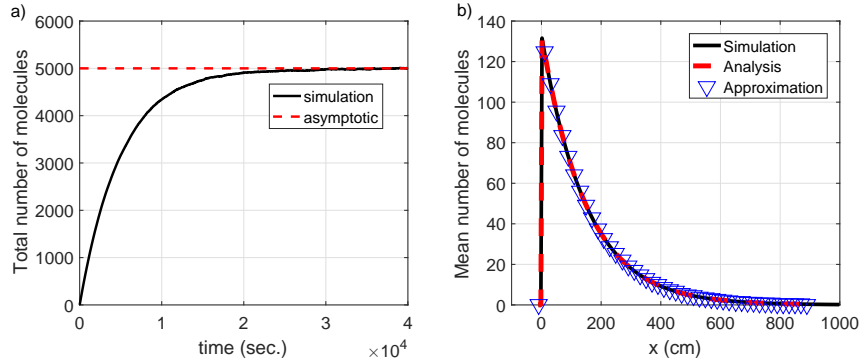}\vspace{-0.5cm}
 \caption{a): Average total number of biomarkers present inside the blood vessel. b): Distribution of the mean number of biomarkers versus $x$. The analysis and approximation results are obtained using (\ref{cx}) and (\ref{velocity_change}), respectively. } 
 \vspace{-0.5cm}
\label{Fig::1}
\end{figure}

Figs. \ref{Fig::1}a) and b) depict the average total number of biomarkers inside the blood vessel over time, along with their distribution with respect to $x$, where the biomarkers are released into the medium at time $t=0$ and position $x=d_0=0$. From Fig. \ref{Fig::1}a) we observe that about $8$ hours (i.e., $28800~$s) after the start of secretion, the total number of biomarkers inside the medium reaches its asymptotic value, which means that the integral in (\ref{final_mean}) has reached its asymptotic value for $t>8~$hours. Furthermore, from Fig. \ref{Fig::1}b) we observe that the biomarkers are exponentially distributed along the $x$-axis for $x>0$, which indicates that the concentration of the biomarkers decreases rapidly as the distance from the release point increases. This motivates cancer detection via MNS as proposed in this paper. Moreover, the concentration of the biomarkers at the left hand side of the release point is approximately zero, which is consistent with the approximation in \eqref{velocity_change}. The results obtained with the exact (analytical) and approximate expressions derived for straight blood vessels in (\ref{cx}) and (\ref{velocity_change}), respectively, are in excellent agreement with the simulation results.

\begin{figure}[!t]
 \centering
 \includegraphics[scale=0.405]{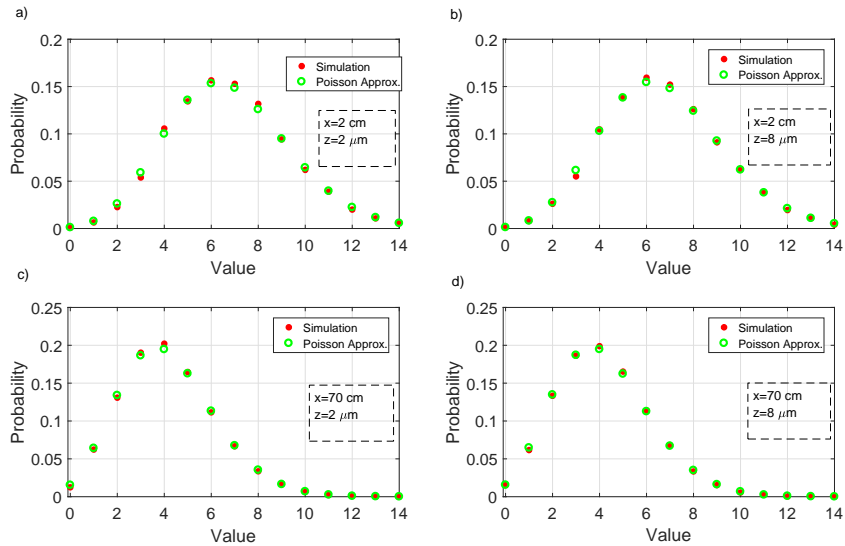}\vspace{-0.5cm}
 \caption{Poisson PMF approximation (\ref{Nbio}) and histogram derived from particle-based simulation for the numbers of biomarkers inside rectangles of area $S=8\times 10^{-4}~$cm$^2$ centered at four different positions inside the blood vessel.}
 \vspace{-0.5cm}
\label{Fig::2} 
\end{figure}

Fig. \ref{Fig::2} shows the histograms for the numbers of biomarkers inside rectangles of area $S= (4~$cm$)\times (2\times 10^{-4}~$cm$) = 8 \times 10^{-4}~$cm$^2$ centered at four different sample positions inside the blood vessel, along with their Poisson PMF approximations according to (\ref{Nbio}). As can be seen, the histogram for the number of biomarkers inside the rectangles is well approximated by a Poisson PMF for all considered cases. This analysis confirms the accuracy of approximating the number of the biomarkers inside a small area by a Poisson PMF.
Finally, by comparing Figs. \ref{Fig::2}a) and \ref{Fig::2}b) as well as Figs. \ref{Fig::2}c) and \ref{Fig::2}d), we observe that the PMF for the number of biomarkers is practically independent of $z$, which validates the assumptions made in Section~III, i.e., a uniform distribution of the biomarkers in $z$-direction for small blood vessels.
%

\begin{figure}[!t]
 \centering
 \includegraphics[scale=0.425]{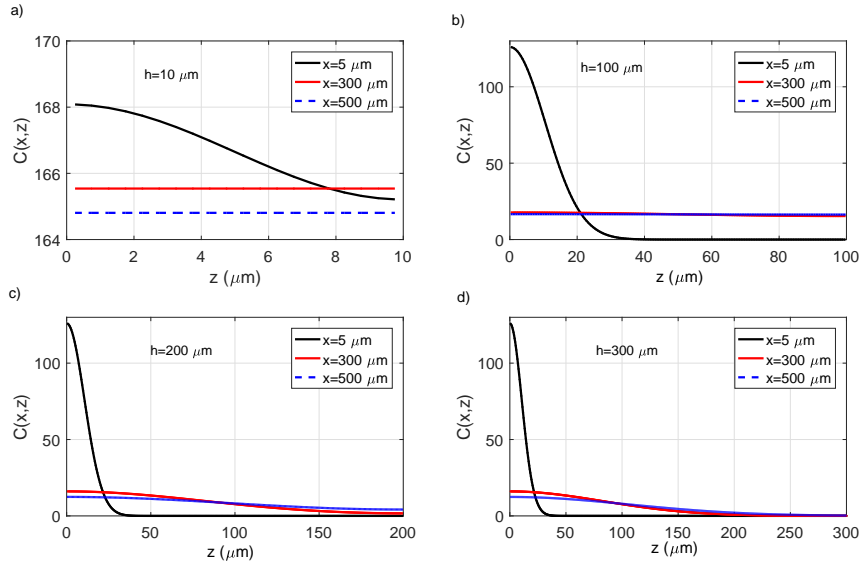}\vspace{-0.5cm}
 \caption{Impact of blood vessel height on the concentration of the biomarkers (\ref{final_mean}) over $z$.} 
 \vspace{-0.5cm}
\label{Fig::4}
\end{figure}

To investigate the effect of the blood vessel height on the distribution of the biomarkers, in Figs. \ref{Fig::4}a), b), c), and d), we plot the biomarker concentration versus $z$ for different values of $x$ and $h$. As can be observed, the distribution of the mean number of biomarkers inside the blood vessel in the vicinity of the secretion point ($x=0$) is not uniform in $z$. However, already for $x=5~\mu$m, for relatively small blood vessels ($h \leq 100~\mu$m), the distribution of the biomarkers with respect to $z$ becomes uniform. On the other hand, for large blood vessels, this uniformity does not hold. As Figs. \ref{Fig::4}c) and \ref{Fig::4}d) show, for the considered values of $x$ and blood vessel heights of $200~\mu$m and $300~\mu$m, the distribution of the biomarkers is not uniform over $z$, but is higher for smaller $z$, as the biomarker secretion point is located at $z=0$. Hence, based on the results obtained in Fig. \ref{Fig::4}, we can conclude that the assumption of approximately uniform concentration for $x\neq d$ is not only valid for capillaries ($h<10~ \mu$m), but for larger blood vessels such as arterioles and venules, with heights smaller than $100~\mu$m.

\begin{figure*}[!tbp]
\centering
\begin{minipage}[t]{0.45\textwidth}\hspace*{-5 mm} 
\centering
\resizebox{1.05\linewidth}{!}{
\includegraphics[scale=0.42]{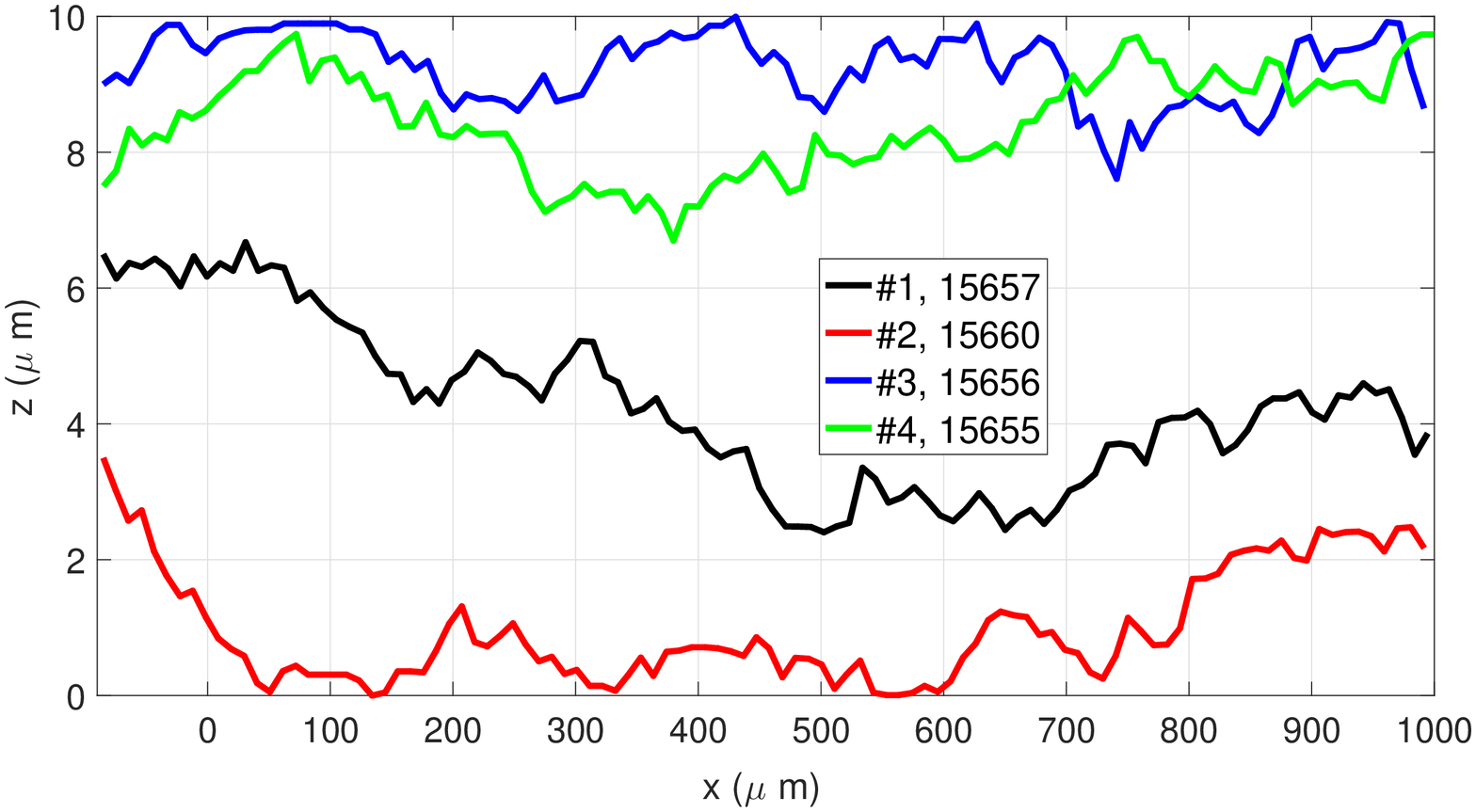}}\vspace*{-6 mm}
\caption{Trajectories of four different MNSs for $\mu = 10^5~$s$^{-1}$, $h= 10~\mu$m, and $d=1$mm. In the legend, the mean value of the MNSs' activation levels (\ref{mean_act_new}) is given.}\vspace*{-3 mm}
\label{Fig::5}
\end{minipage}
\hfill
\begin{minipage}[t]{0.1\textwidth}
\end{minipage}
\vspace*{-0.4 cm}
\begin{minipage}[t]{0.45\textwidth}\hspace*{-5 mm} 
\centering
\resizebox{1.05\linewidth}{!}{
\includegraphics[scale=0.20]{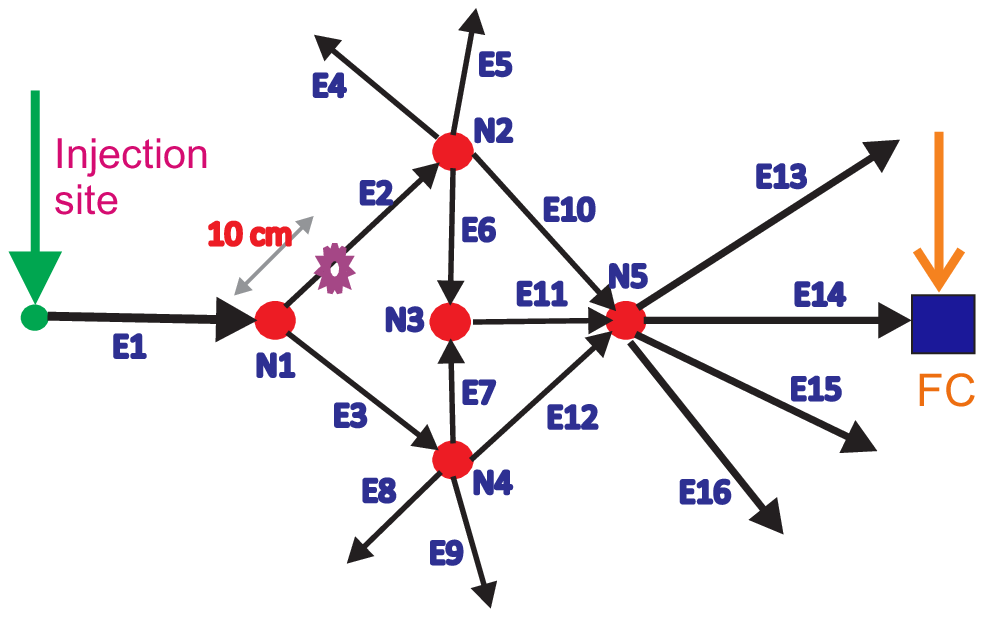}}\vspace*{-5 mm}
\caption{A sample network containing $16$ edges and $5$ nodes.}\vspace*{-3 mm}
\label{Fig.Netw}
\end{minipage}
\end{figure*}

Fig. \ref{Fig::5} depicts the trajectories of four independent MNSs entering a blood vessel of length $d=1000~\mu$m at random $z$ positions. 
In this figure, we investigate the effect of MNS diffusion on the mean value of the activation levels of the MNSs during their passage of the blood vessel. 
We adopt time intervals of length $0.01d/\mathsf{v}$ between consecutive measurements such that each MNS takes approximately $100$ measurements. 
As can be observed, although each MNS diffuses in $z$-direction, the mean values of the activation levels of the MNSs, given in the legend of Fig. \ref{Fig::5}, are relatively close to each other, which is in agreement with the assumption of uniform biomarker distribution over $z$. Moreover, since the number of measurements made by each MNS is $100$ and is equal to the number of measurements that we expected by advection, the effect of diffusion in $x$-direction is negligible in comparison to advection, which justifies the assumption made in Section~\ref{sec:Act} for derivation of (\ref{cond_act}). This is expected since the MNSs' movement by advection, i.e., $d= 10^{-1}~$cm, is two orders of magnitude larger than the typical movement caused by diffusion during the time that the MNS is inside the vessel, i.e., $\sqrt{2D_{\text{n}}d/\mathsf{v}}=2.5\times 10^{-4}~$cm.

\subsubsection{Network of Blood Vessels}

In the following, we study the methods proposed in Section~\ref{Network} for evaluating the biomarker concentration in a network of blood vessels.
%
In Fig. \ref{Fig.Netw}, we show a sample network of blood vessels which is a simplified version of the network given in Fig. \ref{Fig.3}, where all blood vessels have identical cross-section areas. This network contains \emph{five} nodes, $L=16$ edges, and $R=4$ possible routes between the injection site and the FC. For this network, the probability that a released MNS is observed at the FC is $\rho = 0.125$. The MNSs are released at the injection site in edge E$1$ to be finally observed at the FC located in edge E$14$. In general, the lengths of the blood vessels in the human body ranges from less than $1~$cm to $100~$cm \cite{Marieb}. Hence, the distance between the injection site and the FC could be from tens of centimeters to a few meters. In addition, the number of blood vessels inside the network between an injection point and an FC could be on the order of thousands. 
However, to gain insight into the impact of the different parameters for system design, in Fig. \ref{Fig.Netw}, we assume there are only $16$ relevant (straight) edges with relatively large values for their lengths, i.e., $96~$cm for all vessels. We note that the order of the adopted lengths is in line with the values previously used in the MC literature to model blood vessels for drug delivery, cf. \cite{Chahibi}. We also assume equal heights of $h= 80~\mu$m for all edges. The biomarkers are released at edge E$2$, at a distance of $d_0=10~$cm from node N$1$, as depicted in Fig. \ref{Fig.Netw}.
In addition, we consider typical flow velocities for all edges. 
In particular, we assume that the flow velocity in edge E$1$ is $\mathsf{v}_1=0.04~$cm$\cdot$s$^{-1}$. Then, since all edges have the same cross-section area and using (\ref{vel_bif}), the flow velocities in the other edges are obtained as $\mathsf{v}_2=\mathsf{v}_3=0.02~$cm$\cdot$s$^{-1}$, $\mathsf{v}_4=\mathsf{v}_5=\mathsf{v}_6=\mathsf{v}_7=\mathsf{v}_8=\mathsf{v}_9=\mathsf{v}_{10}=\mathsf{v}_{12}=\mathsf{v}_{13}=\mathsf{v}_{14}=\mathsf{v}_{15}=\mathsf{v}_{16}=0.005~$cm$\cdot$s$^{-1}$, and $\mathsf{v}_{11}=0.01~$cm$\cdot$s$^{-1}$. 
Furthermore, in the particle-based simulation, we assume that the velocity changes abruptly at the nodes. Finally, because of the symmetry of the considered network, at each bifurcation node (N$1$, N$2$, and N$5$), the released biomarkers and MNSs enter the outgoing flows with equal probability. For example, each particle that moves through edge E$2$ and arrives at node N$2$, enters each of the edges E$4$, E$5$, E$6$, and E$10$ with probability $0.25$.

\begin{figure}[!t]
 \centering
 \includegraphics[scale=0.43]{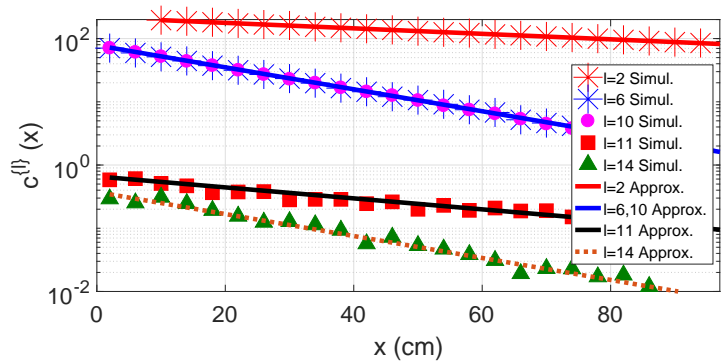}\vspace*{-5.0 mm}
\caption{Concentration of biomarkers, $\CT^{\{l\}}(x)$, in different edges of the network in Fig. \ref{Fig.Netw}. }
\vspace*{-5.0 mm}
\label{Fig::8}
\end{figure}

In Fig. \ref{Fig::8}, the concentration of the biomarkers, secreted by cancer cells, inside the sample network, $\CT^{\{l\}}(x)$, is plotted versus the local $x$-axis of the edges, where we have averaged over $2\times 10^{3}$ independent realizations of the particle-based simulations. 
We have plotted the concentrations of the biomarkers in edges E$2$, E$6$, E$10$, E$11$, and E$14$, which are the only edges that contain biomarkers and are included in at least one route between the injection site and the FC. 
In addition, we also show results for the approximate solution for the concentration of the biomarkers \eqref{appnew__}.
We note that, for the network considered in Fig. \ref{Fig.Netw}, the exact solution for the concentration of the biomarkers \eqref{main_join}, \eqref{bifu} yields practically identical results to the approximate solution. Hence, for clarity, we only show the results for the approximate solution in Fig. \ref{Fig::8}.
As can be observed from Fig. \ref{Fig::8}, the simulated mean number of the biomarkers is in excellent agreement with the approximate result. 
%
In addition, we observe that at junction nodes at which the velocity of the outgoing flow is higher than that of the incoming flow, for instance at node N$3$, two phenomena occur. First, at node N$3$, the concentration at the beginning of the outgoing edge E$11$ is lower than that at the end of edge E$6$. Second, the rate of exponential decrease of the concentration of the biomarkers inside the outgoing edge decreases. That is, inside edge E$6$, the concentration of the biomarkers quickly decreases, while inside edge E$11$, it slowly decreases. Both of these observations are consistent with the expressions provided in (\ref{velocity_change}), (\ref{appnew__}), and (\ref{bifu}), where the concentration is proportional to $(1/{\mathsf{v}})\Ex\left(-\K x/{\mathsf{v}} \right)$.

\subsection{Performance Evaluation}\label{sec:performance}
In the following, we evaluate the performances of the proposed LRT and sum detectors and compare them with a benchmark scheme where the nanosensors are not mobile and are fixed at the FC. 
In particular, the benchmark scheme, which is similar to a conventional blood test, measures the biomarker concentration at a fixed location, and then employs equal gain combining (EGC) for the observations gathered by different sensors. Since the activation levels of fixed nanosensors are Poisson distributed, EGC is the optimal combining rule \cite{Poor_Book_Detection}. 
Furthermore, we assume a large enough time interval of $800~$s between consecutive measurements of the MNSs to have independent measurements on each edge, cf. \cite{Adam}. Hence, the number of measurements taken on the edges having flow velocities of $0.04~$cm$\cdot$s$^{-1}$, $0.02~$cm$\cdot$s$^{-1}$, $0.01~$cm$\cdot$s$^{-1}$, and $0.005~$cm$\cdot$s$^{-1}$ is $3$, $6$, $12$, and $24$, respectively. 
Besides, to have a fair comparison between the proposed scheme and the benchmark scheme, we employ the same number of sensors in both cases.
That is, although the number of MNSs observed at the FC follows the Binomial distribution $\mathsf{Bi}(M_{\text{max}},\rho)$, since $M_{\text{max}}$ MNSs are released at the injection site, we employ $M_{\text{max}}$ nanosensors for the benchmark scheme. 
In addition, we use the same average number of measurements for the MNSs and the fixed nanosensors for the proposed MNS-based approach and the benchmark scheme, respectively. 
Furthermore, MNSs moving through different routes between the injection site and the FC have different activation level means. Applying the concentrations of the biomarkers secreted by cancer cells shown in Fig. \ref{Fig::8} in the inner summation in (\ref{cond_act}), i.e., $m_l \triangleq \sum_{x\in \mathcal{X}^{\{l\}}_{r}} S {\CT^{\{l\}}(x)}/{h}$, we obtain the mean values for the activation levels of the MNSs passing through the different edges as $m_2 = 1.28$, $m_6=m_{10}=0.75$, $m_{11}=0.006$, and $m_{14}=0.003$. We note that for the considered network topology, $m_l$ does not depend on the route that the MNSs take. Also, the mean value of the number of biomarkers secreted by the cancer cells and observed at the FC for each measurement for the benchmark scheme is $10^{-5}$.
Finally, as we are interested in the early stages of a cancer, we assume the presence of a relatively large environmental noise, which reflects the impact of healthy cells. To this end, a Poisson distributed RV with mean $S\xi=0.1$ is employed to model the impact of the environmental noise.

\begin{figure}[!t]
 \centering
 \includegraphics[scale=0.305]{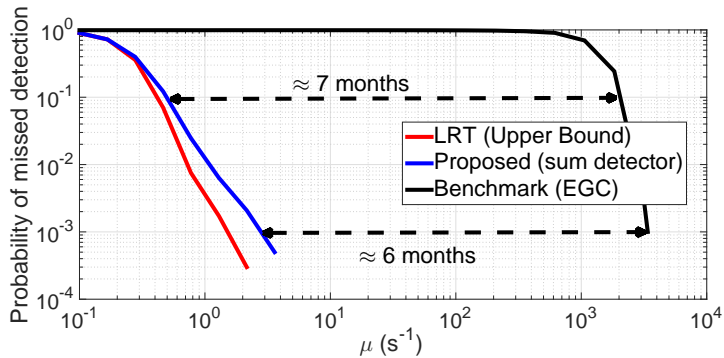}\vspace*{-5.0 mm}
\caption{Probability of missed detection versus the biomarker secretion rate for a given probability of false alarm of $\Pf = \alpha = 10^{-2}$, $M_{\text{max}}=100$, and $\rho = 0.125$.}\vspace*{-5.0 mm}
\label{Fig::12}
\end{figure}

In Fig. \ref{Fig::12}, we compare the performance of the proposed MNS-based approach with that of the benchmark scheme. In particular, in this figure, we plot the probability of missed detection versus the biomarker secretion rate for a given probability of false alarm of $\Pf=\alpha=10^{-2}$, $M_{\text{max}}=100$, and $\rho=0.125$ by employing Monte Carlo simulation for $4\times 10^5$ independent realizations. 
For each secretion rate $\mu$, we have chosen appropriate thresholds for all considered detection schemes such that $\Pf=10^{-2}$, and based on these thresholds we have evaluated $\Pm$. 
In our simulation, for each of the possible routes $\mathcal{R}_1=\{1,2,10,14\}, \mathcal{R}_2=\{1,2,6,11,14\}, \mathcal{R}_3=\{1,3,12,14\}$, and $\mathcal{R}_4=\{1,3,7,11,14\}$, we have evaluated the probability that a released MNS passes through that route ($q_r=1/32, r=1,2,3,4$) as well as the mean number of biomarkers observed by each MNS ($\A_r$ and $\J_r$). Based on this, we have constructed the LRT and sum detectors for the MNS scheme derived in (\ref{mMAP_ns}) and (\ref{practical}), respectively. For the benchmark scheme, we have employed EGC.
%
%
As expected, the LRT detector outperforms the sum detector at the expense of a substantially higher complexity. 
In addition, we can observe that to achieve a probability of missed detection of $\Pm=10^{-3}$ with the proposed MNS-based scheme employing the sum detector, the biomarker release rate has to be around $3~$s$^{-1}$. The benchmark scheme requires release rates larger than $3400~$s$^{-1}$ to achieve the same performance. Fig. \ref{Fig::12} shows that during the time interval in which $\mu$ changes from $0.5~$s$^{-1}$ to $2000~$s$^{-1}$, or equivalently a time interval of \underline{$7$ months}, cf. \eqref{mu}, the benchmark scheme cannot even achieve a probability of missed detection of $\Pm=10^{-1}$. 
For the set of parameters considered here and the considered example network, Fig. \ref{Fig::12} shows that the proposed simple sum detector for the MNS-based approach can detect the presence of cancer at least \underline{$6$~months} earlier than the benchmark scheme. Although the actual CS is much more complex than the sample network in Fig. \ref{Fig.Netw}, we expect that using the proposed MNS-based approach significantly improves the probability of detecting cancer compared to the blood sample tests which are widely used now and are similar to our benchmark scheme.

\begin{figure}[!t]
 \centering
 \includegraphics[scale=0.43]{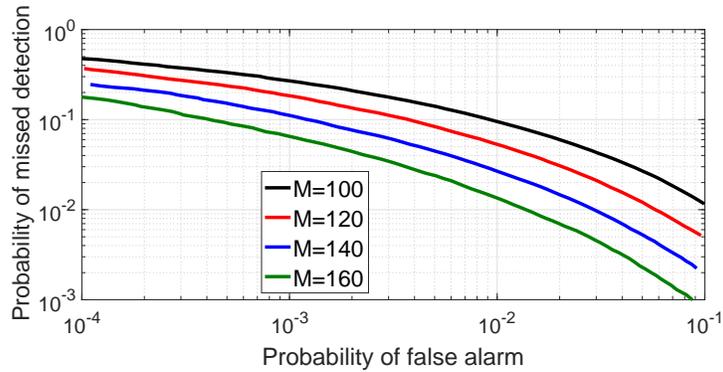}\vspace*{-5.0 mm}
\caption{Probability of missed detection versus probability of false alarm for the proposed sum detector (\ref{practical}) for $\mu=1$s$^{-1}$ and different $M$.}
\label{Fig::13}\vspace*{-5.0 mm}
\end{figure}

Finally, in Fig. \ref{Fig::13}, we investigate the impact of the number of the MNSs observed at the FC on the performance of the proposed sum detector. We show the probability of missed detection versus the probability of false alarm based on Monte Carlo simulation of $4\times 10^5$ realizations. The number of used MNSs is an important design parameter for early cancer detection. As can be observed, as the number of MNSs increases, the performance of the proposed detector improves significantly.

\section{Conclusions}\label{Conc}
In this paper, we studied anomaly detection inside blood vessels where multiple MNSs are injected and pass through the blood vessels. The MNSs sense the presence of an anomaly by detecting the biomarkers secreted by cancer cells. The final decision regarding the presence of anomaly is made at an FC based on the activation levels of the observed MNSs.  
For this collaborative anomaly detection scheme, we first considered a single cancerous blood vessel and derived the spatial distribution of the biomarkers in it as a function of time. Then, we extended the obtained results to a sample network of the CS and analyzed the distribution of the biomarkers in the connected blood vessels of the network.
Based on the biomarker distribution, we modelled the statistics of the activation levels of the MNSs observed at the FC. We also verified the accuracy of the models proposed in this paper via particle-based simulations.
Finally, we derived the optimal decision rule as well as a simple sum detector for the FC and compared their performances via simulation to that of a benchmark scheme. Our simulations revealed that while the optimal LRT detector achieves a higher performance than the sum detector at the expense of a substantially higher complexity, both proposed detectors are superior to the benchmark scheme.
%


\vspace{-0.1cm}


\begin{thebibliography}{60}
\vspace*{-1mm}
\bibitem{MC_Book} T. Nakano, A. Eckford, and T. Haraguchi, \textit{Molecular Communication}. Cambridge Univ. Press, Cambridge, U.K., Sep. 2013.


\bibitem{Akyl_MCNet} I. F. Akyildiz, J. M. Jornet, and M. Pierobon, ``Nanonetworks: A new frontier in communications,'' \emph{Commun. of the ACM}, vol. 54, no. 11, pp. 84--89, July 2011.


\bibitem{Farsad}
N. Farsad, H. B. Yilmaz, A. Eckford, C.-B. Chae, and W. Guo, ``A comprehensive
survey of recent advancements in molecular communication,'' \emph{IEEE Commun. Sur. \and Tut.}, vol. 18, no. 3, pp. 1887--1919, Feb. 2016.


\bibitem{Nakano} Y. Okaie, T. Nakano, T. Hara, and S. Nishio, ``Target detection and tracking by bionanosensor networks,'' Springer Briefs in Computer Science, 2016.

\bibitem{Alok_Mishra} A. Mishra and M. Verma, ``Cancer biomarkers: Are we ready for the prime time?,'' \emph{Cancers}, vol. 2, no. 1, pp. 190--208, Mar. 2010.

\bibitem{LiWu} L. Wuab and X. Qu, ``Cancer biomarker detection: Recent achievements and challenges,'' \emph{Chem. Soc. Rev.}, vol. 44, no. 10, pp. 2963--2997, Mar. 2015.

\bibitem{Keesee} S. K. Keesee, J. V. Briggman, G. Thill, and Y. K. Wu, ``Utilization of nuclear matrix proteins for cancer diagnosis,'' \emph{Crit. Rev. Eukaryot. Gene Expr.}, vol. 6, no. 2-3, pp. 189--214, 1996.

\bibitem{Tarro} G. Tarro, A. Perna, and C. Esposito, ``Early diagnosis of lung cancer by detection of tumor liberated protein,'' \emph{J. Cell. Physiol.}, vol. 203, no. 1, pp. 1--5, Apr. 2005.

\bibitem{Hanash} S. M. Hanash, S. J. Pitteri, and V. M. Faca, ``Mining the plasma proteome for cancer biomarkers,'' \emph{Nature}, vol. 452, pp. 571--579, Apr. 2008.


\bibitem{Miroslava_Stanstna} M. Stastna and J. E. V. Eyk, ``Secreted proteins as a fundamental source for biomarker
discovery,'' \emph{Proteomics.}; vol. 12, no. 4-5, pp. 722--735, Feb. 2012. 

\bibitem{Karagiannis} G. S. Karagiannisa, M. P. Pavloua, and E. P. Diamandisa, ``Cancer secretomics reveal pathophysiological pathways in cancer molecular oncology,'' \emph{Mol. Onc.} vol. 4, no.6, pp. 496--510, Dec. 2010.

\bibitem{Henry} N. L. Henrya and D. F. Hayes, ``Cancer biomarkers,'' \emph{Mol. Onc.}, vol. 6, no. 2, pp. 140--146, Apr. 2012. 

\bibitem{Chen} G. Chen, I. Roy, C. Yang, and P. N. Prasad, ``Nanochemistry and nanomedicine for nanoparticle-based
diagnostics and therapy,'' \emph{ Chem. Rev.}, vol. 116, no. 5, pp. 2826--2885, Jan. 2016.

\bibitem{Drug} U. A. K. Chude-Okonkwo, R. Malekian, B. T. Maharaj, and A. V. Vasilakos, ``Molecular communication and nanonetwork for targeted drug delivery: A survey,'' \emph{IEEE Commun. Surv. Tut.}, vol. 19, no. 4, 4th Quart.,  2017.

\bibitem{Perfezou} M. Perf{\'e}zou, A. Turnerbc, and A. Merkoc, ``Cancer detection using nanoparticle-based sensors,'' \emph{Chem. Soc. Rev.,} vol. 41, no. 7, pp. 2606--2622, Apr. 2012.


\bibitem{Survey_Anomaly}
V. Chandola, A. Baaerjee and V. Kumar, ``Anomaly detection: A survey,'' \emph{ACM Computing Surveys}, Sept. 2009.
\bibitem{Computer_Failor}
V. V. Phoha, \textit{The Springer Internet Security Dictionary}. Springer Verlag, 2002.


\bibitem{Lahouti_Detection}
S. Ghavami and F. Lahouti, ``Abnormality detection in correlated Gaussian molecular nano-networks: Design and analysis,'' \textit{IEEE Trans. NanoBiosci.}, vol. 16, no. 3, pp. 189--202, Mar. 2017.

\bibitem{Reza_TNB} R. Mosayebi, V. Jamali, N. Ghoroghchian, R. Schober, M. Nasiri-Kenari, and M. Mehrabi, ``Cooperative abnormality detection via diffusive molecular communications,'' \emph{IEEE Trans. NanoBiosci.}, vol 16, no. 8, pp. 824--842, Dec. 2017.
\bibitem{Luca} L. Felicetti, M. Femminella, G. Reali, and P. Li{\'o}, ``A molecular communication system in blood vessels for
tumor detection,'' \emph{in Proc. ACM 1st Annu. Int. Conf. Nanosc. Comput. Commun.}, May 2014.
\bibitem{Nakano_Let} S. Iwasaki and T. Nakano, ``Graph-based modeling of mobile molecular communication systems,'' \emph{IEEE Commun. Let.}, vol. 22, no. 2, Feb. 2018.

\bibitem{Brown} P. Brown and C. Palmer, ``The preclinical natural history of serous ovarian cancer: Defining the target
for early detection,'' PLoS Med., vol. 6, no. 7, pp. 1--11, July 2009. 

\bibitem{Poor_Book_Detection}
H. V. Poor, \textit{An introduction to signal detection and estimation theory}, Springer Verlag, New York, 1994.


\bibitem{Martini} F. H. Martini and E. F. Bartholomew, \emph{Essentials of anatomy and Physiology}, Pearson, Dec. 2012.

\bibitem{Hori} S. S. Hori and S. S. Gambhir, ``Mathematical model identifies blood biomarker-based early
cancer detection strategies and limitations,'' \emph{Sci. Transl. Med.}, vol. 3, pp. 109--116, Nov. 2011. 


\bibitem{Malatos} S. Malatos, A. Raptis, and M Xenos, ``Advances in low-dimensional mathematical modeling of the human
cardiovascular system,'' \emph{ J. Hypertens. Manag.}, vol. 2, no. 2, pp. 1--10, Sep. 2016.

\bibitem{Venka} J. Venkatesan, D. S. Sankar, K. Hemalatha, and Y. Yatim, ``Mathematical analysis of Casson fluid model for
blood rheology in stenosed narrow arteries,'' \emph{J. App. Math.}, pp. 1--11, July 2013.



\bibitem{Gentile} F. Gentile, M. Ferrari, and P. Decuzzi, ``The transport of nanoparticles in blood vessels: The effect of vessel 
permeability and blood rheology," \emph{Ann. Bio. Eng.}, vol. 36, no. 2, pp. 25---261, Feb. 2008.

\bibitem{Gill} W. N. Gill and R. Sankarasubramanian, ``Exact analysis of unsteady convective diffusion," \emph{Proc. R. Soc. Lond. A}, vol. 316, no. 1526, pp. 341--350, May 1970.



\bibitem{Atkinson} M. R. Atkinson, M. A. Savageau, J. T. Myers, and A. J. Ninfa, ``Development of genetic circuitry exhibiting
toggle switch or oscillatory behavior in Escherichia coli," \emph{Cell}, vol. 113, no. 5, pp. 597--607, May 2003.

\bibitem{Kramer} B. P. Kramer, A. U. Viretta, M. D. Baba, D. Aubel, W. Weber, and M. Fussenegger, ``An engineered epigenetic transgene switch in mammalian cells,'' \emph{Nature Biotech.}, vol. 22, no. 7, pp. 867--870, July 2004.

\bibitem{Danino} T. Danino, O. Mondrag{\"o}n-Palomino, L. Tsimring, and J. Hasty, ``A synchronized quorum of genetic clocks,'' \emph{Nature}, vol. 463, no. 7279, pp. 326--330, Jan. 2010.

\bibitem{Endres} R. G. Endres and N. S. Wingreen, ``Accuracy of direct gradient sensing by single cells,'' \emph{PNAS}, vol. 105, no. 41, pp. 15749--15754, Oct. 2008.













\bibitem{Rolfe} P. Rolfe, ``Micro- and nanosensors for medical and biological measurements,'' \emph{Sen. Mater.}, vol. 24, no. 6, pp. 275--302, Aug. 2012.


\bibitem{Adam} A. Noel, K. Cheung, and R. Schober, ``Optimal receiver design for diffusive molecular communication with flow and additive noise,'' \emph{IEEE Trans. NanoBiosci.} vol. 13, no. 3, pp. 350--362, Sep. 2014.



\bibitem{Cohen} Y. Cohen and S. Y. Shoushan, ``Magnetic nanoparticles-based diagnostics and theranostics,'' \emph{Curr. Opin. Biotechnol.}, vol 24, no. 4, pp. 672--681, Aug. 2013.

\bibitem{Xie} J. Xie and S. Jon, ``Magnetic nanoparticle-based theranostics,'' \emph{Theranostics}, vol. 2, no. 1, pp. 122--124, Jan. 2012.



\bibitem{Norton} L. Norton, R. Simon, H. D. Brereton, and A. E. Bogden, ``Predicting the course of Gompertzian growth,'' \emph{Nature}, vol. 264, pp. 542--545, Dec. 1976.


\bibitem{Yilmaz_Poiss} 
H.~B. Yilmaz and C.~B. Chae, ``{Arrival modelling for molecular communication via diffusion},'' \emph{Electron. Lett.}, vol.~50, no.~23, pp. 1667--1669, Nov. 2014.


\bibitem{Reza_Receiver} 
R. Mosayebi, H. Arjmandi, A. Gohari, M. Nasiri-Kenari, and U. Mitra, ``Receivers for diffusion-based molecular communication: Exploiting memory and sampling rate,'' \emph{IEEE J. Select. Areas in Commun.}, vol. 32, no. 12, pp. 2368--2380, Dec. 2014.


\bibitem{Siegel} P. Siegel, R. Mose, P. H. Ackerer, and J. Jaffre, ``Solution of the advection-diffusion equation using a combination of discontinuous and mixed finite elements,'' \emph{Int. J. Numer. Methods Fluids},vol. 24, pp. 595--613, Mar. 1997.


\bibitem{Wayan} W. Wicke, A. Ahmadzadeh, V. Jamali, H. Unterweger, C. Alexiou, and R. Schober, ``Molecular communication using magnetic nanoparticles,'' presented in \emph{IEEE WCNC}, Apr. 2018. [Online]. Available: arXiv:1704.04206



\bibitem{Mary} M. P. Wiedeman, ``Dimensions of blood vessels from distributing artery to collecting vein," \emph{Circ. Res.}, vol. 12, pp. 375--378, Apr. 1963.


\bibitem{Mechanics} N. Hwang and R. Houghtalen, ``Fundamentals of hydraulic engineering systems,'' Prentice Hall, Upper Saddle River, NJ. 1996.


\bibitem{Stewart} J. Stewart, \emph{Calculus: Early Transcendentals}, 7th ed., Cengage Learning, 2011.

\bibitem{Roisin} B. Cushman-Roisin, ``Environmental transport and fate,'' Thayer School Eng., Dartmouth College, Univ. Lecture, 2012. [Online]. Available: http://thayer.dartmouth.edu/~d30345d/courses/engs43.html

\bibitem{Binomial} L. Le Cam, ``An approximation theorem for the Poisson binomial distribution,'' \emph{Pac. J. Math.} vol. 10, no. 4, pp. 1181--1197, Nov. 1960.



\bibitem{Marieb} N. E. Marieb and K. Hoehn, \textit{The Cardiovascular System: Blood Vessels}, 9th ed., Human anatomy \& physiology, Pearson Education, 2013. 



\bibitem{Chahibi} Y. Chahibi, M. Pierbon, S. O. Song, and I. F. Akyildiz ``A molecular communication system model for particulate drug delivery systems,'' \emph{IEEE Trans. Bio. Eng.}, vol. 60, no. 12, pp. 3468--3484, Dec. 2013.




\end{thebibliography}
\end{document}